\pageno=1                                      
\input psfig
%
%
%
\font\ninerm=cmr9
\font\eightrm=cmr8
\font\sixrm=cmr6
\font\ninei=cmmi9
\font\eighti=cmmi8
\font\sixi=cmmi6
\skewchar\ninei='177 \skewchar\eighti='177 \skewchar\sixi='177
\font\ninesy=cmsy9
\font\eightsy=cmsy8
\font\sixsy=cmsy6
\skewchar\ninesy='60 \skewchar\eightsy='60 \skewchar\sixsy='60

\font\ninebf=cmbx9
\font\eightbf=cmbx8
\font\sixbf=cmbx6
\font\ninett=cmtt9
\font\eighttt=cmtt8
\hyphenchar\tentt=-1 
\hyphenchar\ninett=-1
\hyphenchar\eighttt=-1
\font\ninesl=cmsl9
\font\eightsl=cmsl8
\font\nineit=cmti9
\font\eightit=cmti8
\newskip\ttglue
\def\tenpoint{\def\rm{\fam0\tenrm}%
  \textfont0=\tenrm \scriptfont0=\sevenrm \scriptscriptfont0=\fiverm
  \textfont1=\teni \scriptfont1=\seveni \scriptscriptfont1=\fivei
  \textfont2=\tensy \scriptfont2=\sevensy \scriptscriptfont2=\fivesy
  \textfont3=\tenex \scriptfont3=\tenex \scriptscriptfont3=\tenex
  \def\it{\fam\itfam\tenit}%
  \textfont\itfam=\tenit
  \def\sl{\fam\slfam\tensl}%
  \textfont\slfam=\tensl
  \def\bf{\fam\bffam\tenbf}%
  \textfont\bffam=\tenbf \scriptfont\bffam=\sevenbf
   \scriptscriptfont\bffam=\fivebf
  \def\tt{\fam\ttfam\tentt}%
  \textfont\ttfam=\tentt
  \tt \ttglue=.5em plus.25em minus.15em
  \normalbaselineskip=12pt
  \let\sc=\eightrm
  \let\big=\tenbig
  \setbox\strutbox=\hbox{\vrule height8.5pt depth3.5pt width0pt}%
  \normalbaselines\rm}
\def\ninepoint{\def\rm{\fam0\ninerm}%
  \textfont0=\ninerm \scriptfont0=\sixrm \scriptscriptfont0=\fiverm
  \textfont1=\ninei \scriptfont1=\sixi \scriptscriptfont1=\fivei
  \textfont2=\ninesy \scriptfont2=\sixsy \scriptscriptfont2=\fivesy
  \textfont3=\tenex \scriptfont3=\tenex \scriptscriptfont3=\tenex
  \def\it{\fam\itfam\nineit}%
  \textfont\itfam=\nineit
  \def\sl{\fam\slfam\ninesl}%
  \textfont\slfam=\ninesl
  \def\bf{\fam\bffam\ninebf}%
  \textfont\bffam=\ninebf \scriptfont\bffam=\sixbf
   \scriptscriptfont\bffam=\fivebf
  \def\tt{\fam\ttfam\ninett}%
  \textfont\ttfam=\ninett
  \tt \ttglue=.5em plus.25em minus.15em
  \normalbaselineskip=10pt 
  \let\sc=\sevenrm
  \let\big=\ninebig
  \setbox\strutbox=\hbox{\vrule height8pt depth3pt width0pt}%
  \normalbaselines\rm}
\def\eightpoint{\def\rm{\fam0\eightrm}%
  \textfont0=\eightrm \scriptfont0=\sixrm \scriptscriptfont0=\fiverm
  \textfont1=\eighti \scriptfont1=\sixi \scriptscriptfont1=\fivei
  \textfont2=\eightsy \scriptfont2=\sixsy \scriptscriptfont2=\fivesy
  \textfont3=\tenex \scriptfont3=\tenex \scriptscriptfont3=\tenex
  \def\it{\fam\itfam\eightit}%
  \textfont\itfam=\eightit
  \def\sl{\fam\slfam\eightsl}%
  \textfont\slfam=\eightsl
  \def\bf{\fam\bffam\eightbf}%
  \textfont\bffam=\eightbf \scriptfont\bffam=\sixbf
   \scriptscriptfont\bffam=\fivebf
  \def\tt{\fam\ttfam\eighttt}%
  \textfont\ttfam=\eighttt
  \tt \ttglue=.5em plus.25em minus.15em
  \normalbaselineskip=9pt
  \let\sc=\sixrm
  \let\big=\eightbig
  \setbox\strutbox=\hbox{\vrule height7pt depth2pt width0pt}%
  \normalbaselines\rm}
%
\def\headtype{\ninepoint}                 
\def\abstracttype{\ninepoint}             
\def\captiontype{\ninepoint}              
\def\footnotetype{\ninepoint}             
\def\refit{\it}                           
\font\chaptitle=cmr10 at 11pt             
\rm                                       

%
%
\parindent=0.25in                         
\parskip=0pt                              
\baselineskip=12pt                        
\hsize=4.25truein                         
\vsize=7.445truein                        
\hoffset=1in                              
\voffset=-0.5in                           

\newskip\sectionskipamount                
\newskip\aftermainskipamount              
\newskip\subsecskipamount                 
\newskip\firstpageskipamount              
\newskip\capskipamount                    
\newskip\ackskipamount                    
\sectionskipamount=0.2in plus 0.09in
\aftermainskipamount=6pt plus 6pt         
\subsecskipamount=0.1in plus 0.04in
\firstpageskipamount=3pc
\capskipamount=0.1in
\ackskipamount=0.15in
\def\sectionskip{\vskip\sectionskipamount}
\def\aftermainskip{\vskip\aftermainskipamount}
\def\subsecskip{\vskip\subsecskipamount} 
\def\firstpageskip{\vskip\firstpageskipamount}

%
%
\nopagenumbers                            
\newcount\firstpageno                     
\firstpageno=\pageno                      
\newcount\chapno                          

\def\rightheadline{\headtype\phantom{\folio}\hfil\runningtitletext\hfil\folio}
\def\leftheadline{\headtype\folio\hfil\runningauthortext\hfil\phantom{\folio}}
\headline={\ifnum\pageno=\firstpageno\hfil
           \else
              \ifdim\ht\topins=\vsize           
                 \ifdim\dp\topins=1sp \hfil     
                 \else
                     \ifodd\pageno\rightheadline\else\leftheadline\fi
                 \fi
              \else
                 \ifodd\pageno\rightheadline\else\leftheadline\fi
              \fi
           \fi}

\def\bottomnumber{\hss\tenrm[\folio]\hss}
\footline={\ifnum\pageno=\firstpageno\bottomnumber\else\hfil\fi}

%
%
%
%
\outer\def\mainsection#1
    {\vskip 0pt plus\smallskipamount\sectionskip
     \message{#1}\vbox{\noindent{\bf#1}}\nobreak\aftermainskip\noindent}
 
\outer\def\subsection#1
    {\vskip 0pt plus\smallskipamount\subsecskip
     \message{#1}\vbox{\noindent{\bf#1}}\nobreak\smallskip\nobreak\noindent}
 
\def\backup{\nobreak\vskip-\baselineskip\nobreak\vskip-\subsecskipamount\nobreak
}

\def\title#1{{\chaptitle\leftline{#1}}}
\def\name#1{\leftline{#1}}
\def\affiliation#1{\leftline{\it #1}}
\def\abstract#1{{\abstracttype \noindent #1 \smallskip\vskip .1in}}
\def\ref{\noindent \parshape2 0truein 4.25truein 0.25truein 4truein}
\def\caption{\noindent \captiontype
             \parshape=2 0truein 4.25truein .125truein 4.125truein}

\def\footnote#1{\edef\fspafac{\spacefactor\the\spacefactor}#1\fspafac
      \insert\footins\bgroup\footnotetype
      \interlinepenalty100 \let\par=\endgraf
        \leftskip=0pt \rightskip=0pt
        \splittopskip=10pt plus 1pt minus 1pt \floatingpenalty=20000
        \textindent{#1}\bgroup\strut\aftergroup\strut\egroup\let\next}
\skip\footins=12pt plus 2pt minus 4pt 
\dimen\footins=30pc 

%
%

\def\@{\spacefactor 1000}

\def\,{\pcomma} 
\def\pcomma{\relax\ifmmode\mskip\thinmuskip\else\thinspace\fi}

\def\oversim#1#2{\lower0.5ex\vbox{\baselineskip=0pt\lineskip=0.2ex
     \ialign{$\mathsurround=0pt #1\hfil##\hfil$\crcr#2\crcr\sim\crcr}}}
\def\simgt{\mathrel{\mathpalette\oversim>}}

\def\runningtitletext{Formation of Stellar Clusters}
\def\runningauthortext{C. J. Clarke et. al.}

\def\etal{{\rm et al.~}}
\def\simless{\mathbin{\lower 3pt\hbox
   {$\rlap{\raise 5pt\hbox{$\char'074$}}\mathchar"7218$}}}   
\def\simgreat{\mathbin{\lower 3pt\hbox
   {$\rlap{\raise 5pt\hbox{$\char'076$}}\mathchar"7218$}}}   
\def\etal{{\rm et al.}}

\null
\firstpageskip

{\baselineskip=14pt
\title{The Formation of Stellar Clusters}
}

\vskip .3truein
\name{Cathie J. Clarke, Ian A. Bonnell }
\affiliation{Institute of Astronomy, Madingley Road, Cambridge, CB3 0HA, UK}
\vskip .2truein
\leftline{and}
\vskip .1truein
\name{Lynne A. Hillenbrand}
\affiliation{California Institute of Technology; Pasadena, CA 91125, USA}
\vskip .3truein


\abstract{We review recent work that investigates the formation of stellar
clusters, ranging in scale from globular clusters through open
clusters to the small scale aggregates of stars observed in T
associations. In all cases, recent advances in understanding have been
achieved through the use of state of the art stellar dynamical and gas
dynamical calculations, combined with the possibility of
intercomparison with an increasingly large dataset on young
clusters. Among the subjects that are highlighted are the frequency of
cluster-mode star formation, the possible relationship between
cluster density and the highest stellar mass, subclustering and the dynamical
interactions that occur in compact aggregates, such as binary star
formation.  We also consider how the spectrum of stellar masses
may be shaped by the process of competitive accretion in dense
environments and  how cluster properties, such as mass
segregation and cluster morphology, can be used in conjunction with
numerical simulations to investigate the initial conditions for
cluster formation. Lastly, we contrast bottom-up and top-down scenarios
for cluster formation and discuss their applicability to the formation
of clusters
on a range of scales. 
}


\mainsection{I.~~Introduction}
\backup

Observations indicate that stars frequently form in clustered
environments -- in rich clusters of many hundreds to many thousands of
stars, or in smaller groups and aggregates containing of order ten to
a few tens of stars.  It is only recently, however, that the
properties of young clusters are beginning to be well characterised.
Cluster formation is important, therefore, insofar as it is a
fundamental unit of star formation. It is also becoming increasingly
apparent, given the high stellar densities measured in young clusters
and therefore the possible role of encounters, that whether a star
forms in a cluster or in isolation may be important in determining its
fundamental properties, such as its mass, binarity  or possession
of planets.

This Chapter will concentrate on the issue of how observed young clusters
can be used to deduce the conditions in
clusters at birth. In particular, it stresses the interplay between 
observations and numerical simulations, which allows one to 
address a number of questions
regarding the initial shapes, mass distributions and dynamical states of 
clusters, as well as exploring how likely it is for clusters to
survive as bound structures. Significant observational and computational
advances in recent years make this exercise particularly timely.  
On the observational front, deep wide-field imaging at infrared wavelengths
and multi-fibre spectroscopy have brought a wealth of data
concerning the state of clusters at increasingly young ages. Numerical
simulations have also advanced considerably through the development of
hydrodynamic codes that can deal with the highly inhomogeneous
conditions in star forming gas. Of particular significance is the
recent advent of special purpose `GRAPE' hardware for the calculation
of gravitational forces (Okumura et al. 1993). This innovation has
heralded a new era in Nbody calculations, so that it is now straightforward
to perform simulations (over tens of dynamical times) in which the
number of particles matches the number of stars, even in the case of
populous clusters containing many tens of thousands of stars. 

The reason it is desirable to derive the basic characteristics of
clusters at birth is because of the light such information sheds on
{\it how} clusters form.  Observational constraints on the age spread
in clusters, the time sequence of star formation as a function of
stellar mass, and the degree of subclustering are all important
constraints on theoretical models.  We defer a fuller discussion of
current theoretical ideas until Section VII, but here indicate some of
the issues in order to motivate the intervening Sections of the paper.

Historically, cluster formation theories considered the monolithic
top-down collapse of Jeans unstable gas, and the main issue therefore
concerned the number of fragments (`stars') formed during collapse
(Hoyle~1953; Larson~1978).  Such studies envisaged
rather smooth initial conditions and therefore interest focused on the
amplification of initially linear density perturbations and on the
efficiency of cooling during collapse.  Two facts about the state of
star forming gas in molecular clouds however render this picture
obsolete. Firstly, the thermal energy content of the gas is negligible
compared with the energy density in (assumed MHD) turbulence: hence
the question of how pieces of the cloud collapse to form stars does
not hinge primarily on cooling but instead on their ability to
decouple from the magnetic field. Secondly, molecular clouds are
extremely inhomogeneous (e.g. Vazquez Semadeni \etal, this volume),
consisting of a floculent ensemble of structures within structures
(for a hierarchical description of star forming clouds, see Chapter by
Elmegreen, Zinnecker, Pudritz and Efremov this volume).

This inhomogeneity of the parent gas has several implications for cluster
formation. For one thing,  it renders trivial the question: `
Why are stars clustered
at birth?', since at some level this reflects the structure of the
star forming gas, albeit modified by dynamical effects (see Klessen,
Burkert and Bate 1998 for a first attempt to model cluster formation
from highly inhomogeneous initial conditions). 
It should be noted in passing that the fractal dimension characterising
the distribution of young stars is {\it not} equal to that of the gas, implying
either 
that the star formation process engenders tighter clustering or else
that stars form from the most tightly clustered component of the molecular
gas (Larson 1995). 

The complex density structure of star forming clouds also raises
questions as to the degree of coordination that is required to form a
cluster. It is well known (e.g. Lada et al. 1984; Goodwin 1997; see
Section V below) that the formation of a {\it bound} cluster requires
that a high fraction (30-50 \%) of the gas must be turned into stars
before destructive feedback mechanisms from massive stars come into
play: in practice this means a high conversion efficiency within a few
cluster dynamical times.  Such locally coordinated star formation is a
natural expectation in top-down scenarios (i.e. where the structure
develops as a result of gravitational instabilities during collapse).
It is not however the obvious outcome if star formation is taking
place in an already highly structured environment, unless some
external agent can synchronise the onset of star formation in a set of
discrete, and mutually independent, clumps.  Such triggered star
formation is therefore an attractive possibility theoretically, and
there are clear examples (such as in IC~1396, Patel \etal 1998; the
Rosette Molecular Cloud, White \etal 1997; IC~1805, Heyer \etal 1996;
Gem OB1, Carpenter \etal 1995; and in more isolated ``bright rim''
regions, Sugitani \etal 1991,1994) where the location of young stellar
objects -- and clusters -- in the dense gas swept up by expanding HII
regions lends credence to this scenario (Elmegreen and Lada 1977). In
other cases, however, the locations of young clusters give no hint of
external triggering (e.g. Taurus, NGC~2264).  Thus a key question
(whether cluster formation is induced or spontaneous) remains
unanswered at the present time.  Clearly, the derivation of cluster
parameters at birth (particularly the age spread of stars within a
cluster and the initial degree of sub-clustering) can shed
considerable light on this question.

\mainsection{{I}{I}.~~Observations of Young Clusters}
\backup

Clusters are useful laboratories for star formation studies since they
provide stellar samples of constant metallicity at
approximately uniform distance. The task of identifying and characterizing
clusters so young that they are still embedded in the molecular material
from which they formed has been considerably aided within the past decade
by near-infrared imaging capabilities.  Near-infrared surveys penetrate
through an order of magnitude more column density than does visual imaging
and allow us to see clusters closer to the epoch of their formation
(e.g. Figure~1 shows the  infrared H-band image of Mon R2, Carpenter
\etal~1997, see also Plate~1).

\vskip.05truein
\vbox{{\hbox{\centerline{\psfig{figure=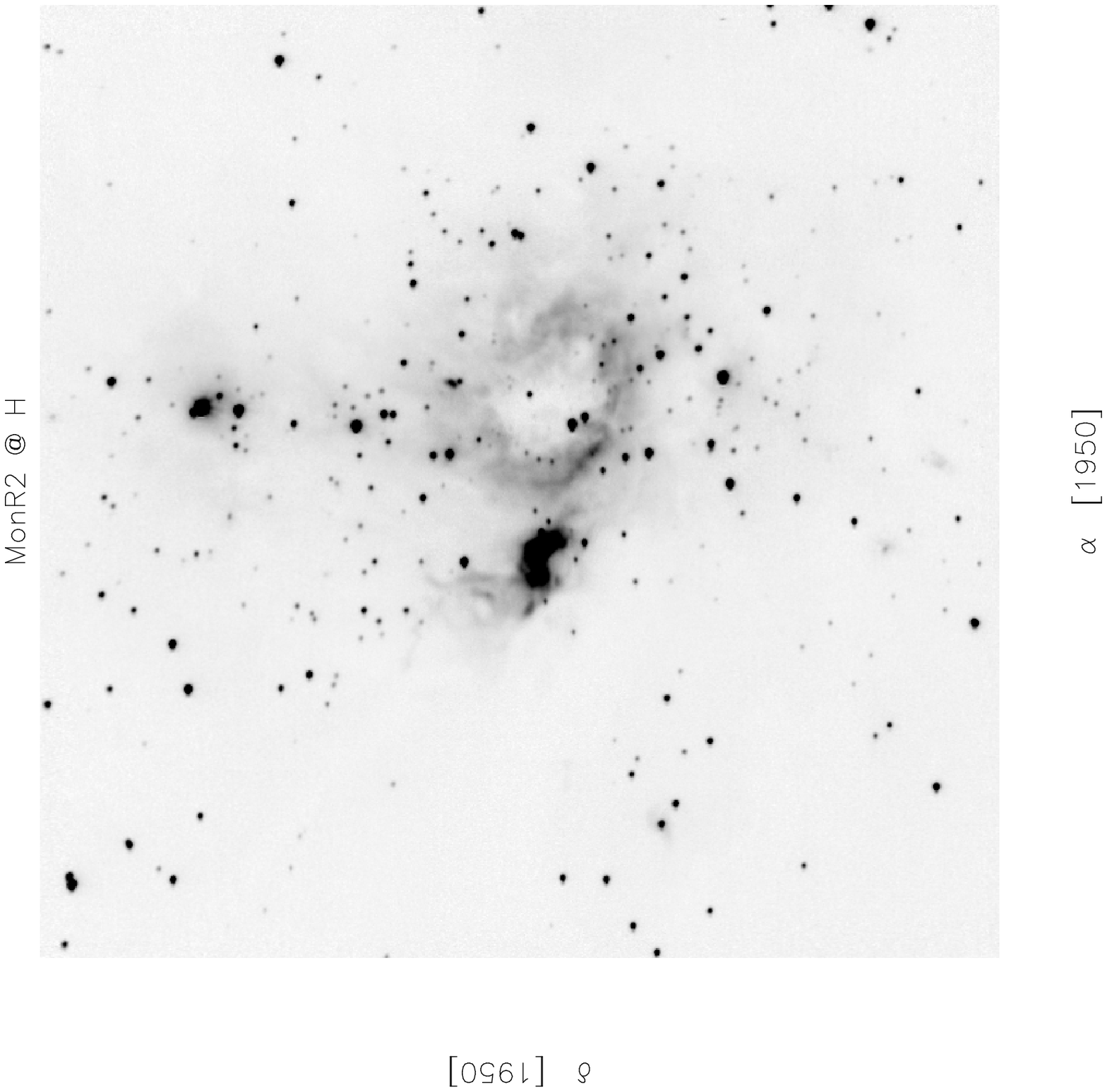,width=2.5truein,height=2.7truein,rwidth=2.6truein,rheight=2.7truein}}}\noindent {\bf Figure 1.}  The Mon R2 cluster imaged in the near-infrared H-band 
(from Carpenter \etal~1997).  The field of view is $\sim 3.2 \times 3.2$ arcmin$^2$
corresponding to $\sim 0.8 \times 0.8$ pc$^2$ for a distance of 830 pc.   
The central cluster is completely embedded
and contains $>$300 stars within a 0.4 pc diameter.
}\vskip.15truein}

In what follows we consider only young stellar populations located within 1kpc 
of the Sun and focus on infrared surveys, as summarized in Table 1.
We distinguish between biased surveys or deep imaging of some interesting class
of object - e.g. molecular outflows, IRAS point sources, 
Bok globules, OB or Herbig Ae/Be stars -
and the often shallower, unbiased surveys of large regions containing 
molecular material.  We highlight two issues: the relative
importance of isolated versus cluster mode star formation, and the apparent
association of high mass star formation with the formation of clusters.
First, however, we briefly touch on some of the problems involved in the
identification and characterisation of clusters.

Clusters are usually identified via enhanced surface
density relative to the background.  An obvious disadvantage of this method 
is that since a given cluster subtends a smaller angle at
larger distances, distant clusters are more readily identified, 
although this effect is partially offset by diminishing survey
sensitivity at larger distances.  Further problems are the correct
subtraction of foreground and background sources, 
and the tendency for patchy absorption to act as
a source of spurious clustering. As cluster surveys are extended to
regions that are increasingly embedded (i.e. closer to the $``t=0''$ of
star formation), it is becoming increasingly necessary to interpret
clustering statistics in conjunction with molecular extinction maps.
These not only allow one to  distinguish between true clustering and
the apparent clustering of sources in windows of low extinction, but also allow
more accurate subtraction of foreground and background sources.
It should be stressed that in what follows, the term cluster is used to describe
apparent groupings of stars in projection; since kinematic data is not usually
available, it is not possible to make the conventional distinction
between clusters and associations on the basis of whether or not
they are gravitationally bound.
It should also be noted that the detection of clustering in molecular clouds
is strongly affected by the age of the system.  With velocity dispersions
of 1-2 kms$^{-1}$, smaller and less dense clusters can disperse quickly,
possibly causing us to have over-estimated  ``typical'' cluster 
membership numbers and projected densities.

 The first large scale near-infrared imaging survey of a molecular
cloud is the oft-quoted work of Lada et al. 1991b, which covered over
50 pc$^2$ of the Orion B molecular cloud (see also Li et al. 1997).
Subsequently, similar unbiased surveys have been conducted in a number
of other star forming regions: the Orion A cloud (Strom,Strom and Merrill
1993; Jones et al. 1994; Ali and DePoy 1995), 
NGC 2264 (Piche 1993; Lada et al. 1993; Strom et al. 1999), 
IC 348 (Lada and Lada 1995), NGC 1333 (Aspin and Sandell 1994,1997;
Lada, Alves and Lada 1996), the Rosette Molecular Cloud (Phelps and Lada
1997), R CrA (Wilking et al. 1997), Taurus (Itoh et al. 1996), 
and the most thoroughly studied region, Ophiuchus (Rieke et al. 1989; 
Barsony et al. 1989; Greene and Young 1992; Comeron et al. 1993;
Strom, Kepner and Strom 1995; Barsony et al. 1997). Clusters are
found in all cases, and generally there is an accompanying  distributed 
population of young stars as well.


It is obviously of interest to assess what fraction of stars form in
clusters. The strong clustering of massive stars has been evident for a long
time (e.g. Ambartsumian 1947; Blaauw 1964), but it is the
advent of near-infrared imaging (as reviewed by Zinnecker et al. 1993) which has revealed that low mass stars form 
abundantly in the vicinity of high mass stars
and thus share in the cluster environment at birth. 
The results of
unbiased surveys of star forming regions suggest that the fraction of star
formation taking place in clusters varies quite strongly from place to place.
This is particularly striking in the case of the Orion giant molecular cloud
where marked differences are found between the A and B clouds  
(see Meyer and Lada, 1998 for a fuller discussion). 
In the Orion B cloud, almost all (96\%) of associated infrared sources 
are thought to be in clusters.  In the Orion A cloud, by contrast, 
there is a significant distributed population with only 50-80\% of the stellar
population formed in clusters (the range depending on whether one
does not or does count the Orion Nebula Cluster (henceforth the ONC)).
The Orion A result is more consistent with what has been found in other surveys
of molecular clouds, where the fraction of stars located in projected
density enhancements (``clusters'') is 50-70\% 
(Taurus, Gomez et al. 1993; NGC~2264, Piche et al. 1993;
NGC~1333, Lada et al. 1996; IC~348, Lada et al. 1995).
We note, however, that while large fractions of the most dense and ``active'' 
areas of many clouds have been surveyed in the near-infrared, in no case
has the entirety of any giant molecular cloud been mapped. 
Thus the fraction
of stars observed to have formed in and out of clusters and aggregates is 
still uncertain.
Significant progress on characterising the cluster forming
properties of different regions is likely to come from analysis of
data on star-forming regions contained in the near-infrared all-sky
surveys (2MASS,DENIS).

Although it is not clear why the fraction of star formation taking
place in dense clusters should vary from cloud to cloud, all of the
regions surveyed thus far seem to support a basic picture in which
{\it the majority of star formation at all masses takes place in
clusters}. 

A similar picture in which clustering is a common, but not ubiquitous,
accompaniment to star formation emerges from the biased surveys of localized
regions associated with some indicator of very recent star formation. 
In L1641, 25\% of the young IRAS sources surveyed by Chen and Tokunaga (1994)
were found to have near-infrared clusters.  From the same survey, 63\% of 
the outflow regions contain clusters while in a broader survey Hodapp (1994) 
found 33\% of molecular outflow sources to have clusters.
Of 44 bright-rimmed clouds (regions thought to be examples of triggered star formation)
containing IRAS sources surveyed by Sugitani \etal (1995) ``most'' are claimed
to harbor small clusters.  On the other hand, Carballo and Sahu (1994)
found no evidence for clustering around the IRAS sources in their
survey and a similar null result was obtained from deep imaging of
Bok globules (Yun and Clemens 1994).

An even higher incidence of clustering appears in the surveys
of regions containing massive stars.  For example, the unbiased surveys of 
Orion B reveal that clusters are associated only with the bright stars exciting the
conspicuous nebulae in the region; stated in reverse, each of the high-mass
stars in Orion B is accompanied by a cluster. A similar connection is suggested
from the biased surveys.  The highest incidence of clustering (19/20 cases) 
occurred in the survey of outer Galaxy IRAS sources radio-selected 
to contain OB stars  (Carpenter et al. 1993).  Similarly, the near-infrared surveys 
of Herbig Ae/Be stars by Hillenbrand (1995) and Testi et al. (1997, 1998) 
(see also Aspin and Barsony 1994; Wilking et al. 1997), 
indicate that clusters are present
around those Ae/Be stars with masses in excess of $3-5 M_{\odot}$, with
little evidence of clustering around less massive objects (see also Chapter
by Stahler et al.). 

One possible correlation in the data is that between stellar density and the 
mass of the most massive cluster member 
(Hillenbrand 1995; Testi et al. 1999; see Figure 2). 
Since clusters exhibit a rather small range of projected radii
(see Table 1 and also Fig. 1 in Testi et al. 1999)  
this also translates into a correlation between
the cluster membership number $N$ and most massive star.
It is at present unclear whether this correlation represents a
genuine physical requirement of high density or $N$ for massive
star formation (see, e.g., Bonnell, Bate and Zinnecker 1998), or whether
it is merely a consequence of random drawing from an IMF, which
would imply that a given cluster is more likely to contain a massive
star if it has high $N$. Distinguishing between these
two possibilities will require a significant number of small-N clusters
to compare with the mass-distributions in large-N clusters.

\vskip.05truein
\vbox{{\hbox{
{\psfig{figure=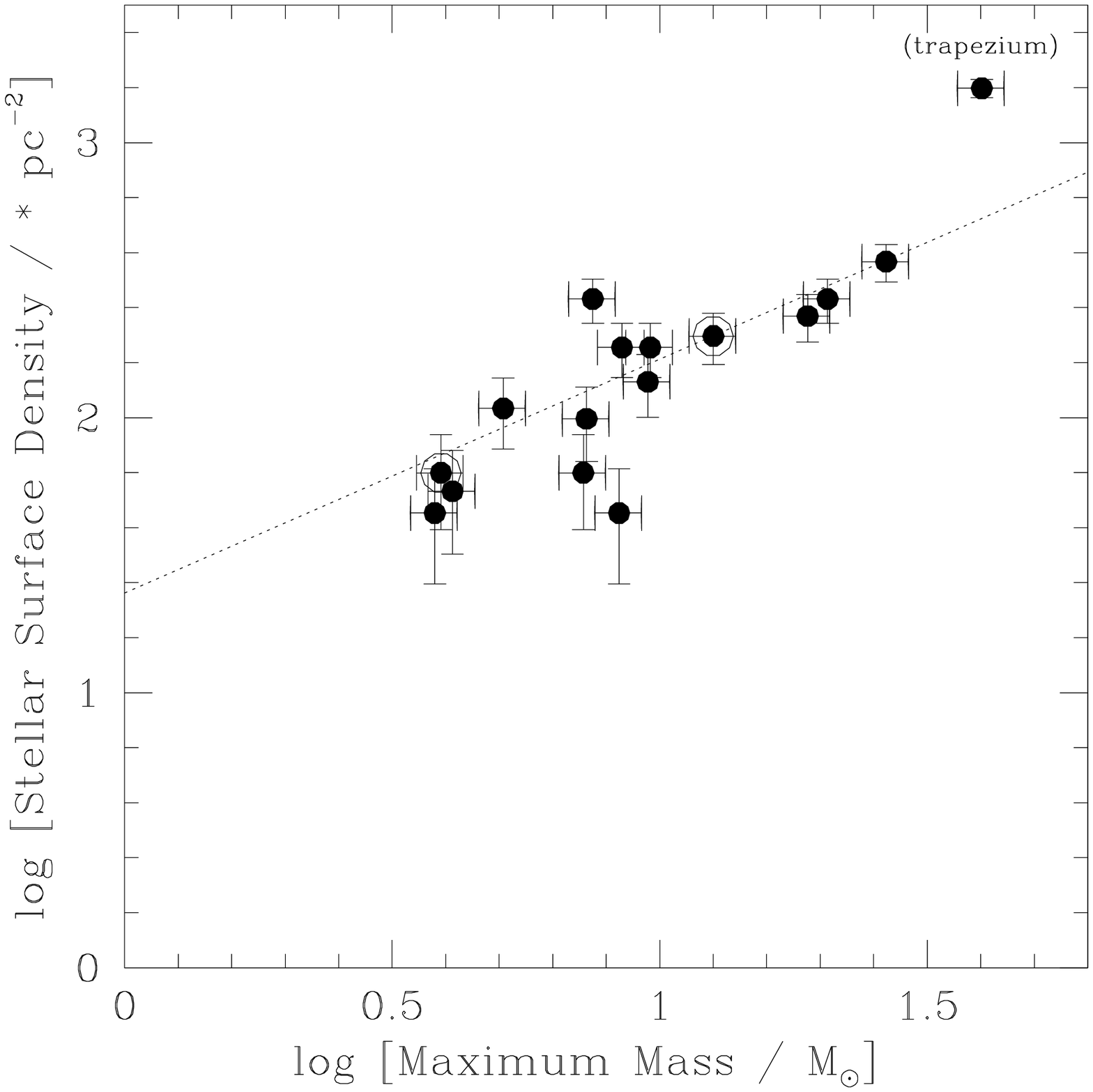,width=2.15truein,height=2.15truein}}
{\psfig{figure=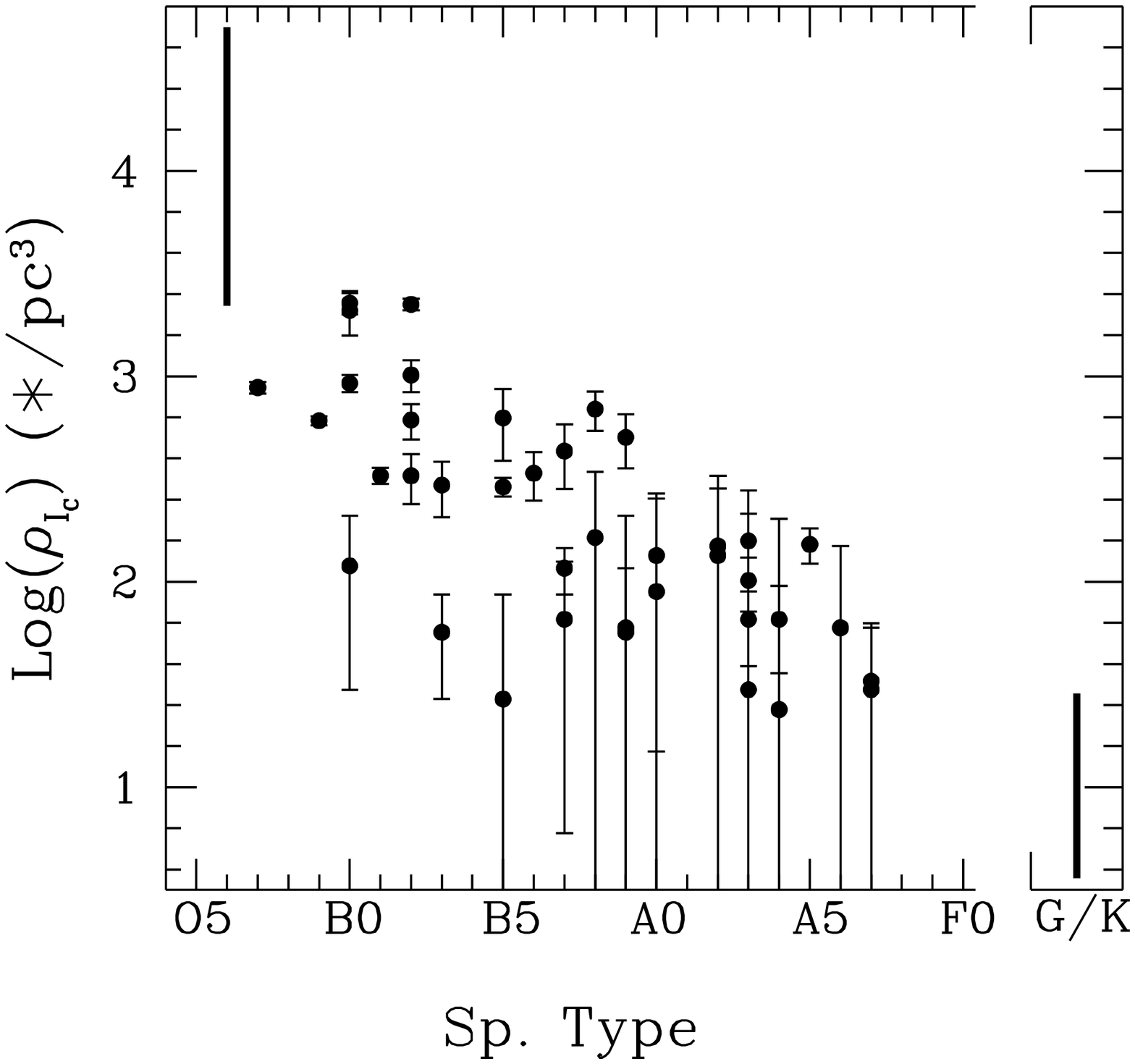,width=2.15truein,height=2.15truein}}
}\noindent {\bf Figure 2.} 
Quantification of clustering around Herbig Ae/Be stars.
In the left panel, stellar surface density (pc$^{-2}$)
is plotted  against mass (M$_\odot$) of the most massive star
(from Hillenbrand 1995); random errors of 10\% in mass and
$\sqrt{N}$ in star counts at K-band are shown, along with a least-squares fit.
In the right panel, stellar volume density (pc$^{-3}$)
is plotted against spectral type of the Ae/Be star (from Testi \etal~1998);
counting statistics in the I-band source counts are shown.
For scaling reference only the innermost region of the Orion Nebula Cluster
is plotted in the upper right of panel (a) and the upper left of panel (b).
Regions containing more than one Ae/Be star do not occupy any preferred
location in these diagrams. }\vskip.15truein}

A strong association is found between the location of clusters and
of dense, massive molecular cloud cores. For example, all the clusters in
Orion B are associated with CS cores (Lada et al. 1991a), as are those
in Orion A (Strom, Strom and Merrill 1993). Moreover, those CS cores
in L1630 which are associated with clusters contain a higher fraction
of very dense gas ($ \simgt 10^5$ cm$^{-3}$) than the clusterless cores
(Lada, Evans and Falgarone 1997). Likewise, in
the Rosette molecular cloud, the seven embedded clusters discovered by
Phelps and Lada (1997) are all associated with moderately massive
molecular cores (as traced by $^{13}$CO), although the majority
of massive cores do not harbour clusters. 
These results may suggest that gas density, as opposed
to mass, may be the critical factor in promoting cluster formation,
although follow-up studies in a density sensitive tracer such as
CS or NH$_3$ are required in the Rosette region to confirm the hypothesis.

Cluster parameters as summarized in Table 1 are not directly
comparable between the various regions due to inconsistencies in the analyses. 
In particular, we emphasize that the values given for the number of stars,
and hence the number density, are in all cases likely to be lower limits.
To effect a rigorous comparison of the stellar populations
emerging from molecular clouds, we ideally need surveys to uniform completeness in mass
(to $<0.1M_\odot$) over a known range in age ($\sim$3 Myr)
and through some given value of the extinction (10-20 mag, say).
However, if one assumes from current databases that cluster sizes 
are good to a factor of $2$ and cluster densities
are good to a factor of $3-5$, intercomparisons can be made. 

The sizes of young (ages less than a few $\times 10^6$ years) clusters
appear fairly uniform (in the range $0.2-0.8$ pc FWHM)
and, notably, are a factor of $5-10$ times smaller 
than the typical sizes of Galactic open clusters 
(with ages a few $\times 10^7$ to $10^9$ years; 
Phelps and Janes 1994; Janes, Tilley and Lynga 1988). Several young clusters
within a kiloparsec of the Sun are sufficiently populous to rank as
candidate proto-open clusters,
although it is uncertain that they will remain bound once their
component gas is removed (see
Section V).
Note that we exclude OB associations from Table 1 and stress that these are
considerably bigger (a few tens of pc). 
Cluster densities have a spread which is larger than the errors, and 
span a few $\times 10^2$ to a few $\times 10^4$ stars pc$^{-3}$,
the latter value corresponding to the core of the ONC. Such
densities correspond to volume averaged values in the range several $10^3$
to several $10^5$ molecules cm $^{-3}$, consistent with the strong correlation
between clusters and concentrations of {\it dense} molecular gas. 

Finally, we turn from a description of the gross parameters of young
clusters to a brief mention of recent attempts to characterise their
stellar content in detail.  This exercise involves combined
spectroscopy and photometry in order that stars can be individually
de-reddened and placed in theoretical HR diagrams where their location
can in principle (i.e.  given well determined theoretical tracks) be
used to determine stellar masses and ages (cf. Hillenbrand 1997;
Herbig 1998; Strom et al. 1999).  Recently, this traditionally
optical technique has been successfully applied in the near-infrared
to study deeply embedded populations -- those obscured by 10-50
magnitudes of interstellar and circumstellar extinction (cf. Hodapp
and Deane 1993; Greene and Meyer 1995; Carpenter \etal 1997; Hanson
\etal 1997; Luhman and Rieke 1998; Meyer \etal 1999).  The
observationally intensive nature of this exercise means that few
clusters have been studied in detail as yet.  Clearly, the information
yielded on the mass distribution (cf. Meyer \etal~1998, this volume)
and age spread of stars in clusters can be expected to have a major
impact on cluster formation theories in the next few years.  It is
notable, for instance, that the relatively old cluster IC~348 appears to 
show evidence for ongoing star formation over a considerable period
(up to $10$ Myr; 
Lada and Lada~1995; Preibisch, Herbig and Zinnecker 1997; Herbig~1998) whereas a high fraction of mass in the
ONC would seem to have  been converted into stars in less
than a million years (Hillenbrand~1997).

\mainsection {{I}{I}{I} ~~Clusters within clusters?} 
\backup

Images of young clusters often contain substructure that is readily
identifiable by eye. Examples occur on a wide range of size scales: at
one extreme, the `Super Star Clusters' (SSCs) observed in interacting
galaxies such as the Antennae (Whitmore et al. 1998) comprise ensembles of 
tens to hundreds of clusters within a couple of hundred pc,
while the SSCs themselves appear to be
clustered in groups of a few. Nearby star forming regions also contain a
wealth of sub-structure (see, for example, Gomez et al. 1993; and
Chapter by Elmegreen et al., this volume).

The issue of sub-clustering of stars at birth is a fundamental one because
it defines the local potential in which stars form and determines whether
or not interactions between adjacent protostars (and associated gas/discs)
play an important role in the star formation process. 
Compact clusters with few members are however short-lived
against dynamical dissolution (see Section IV), so that by the age at
which clusters are observed in a relatively unobscured state (generally
a million years or so) much of the original sub-structure may have been
erased, although traces may remain in positional and velocity data. The link
between the structure of observed clusters and the structure that they had
at birth therefore needs to be mapped out via numerical simulations
(see, for example, Goodwin 1997 in the context of the LMC globular clusters).

Apart from these questions of how observed structure relates to
structure at birth (which can be addressed by simulations) there is
the equally important issue of how the statistical significance of
apparent substructure is to be assessed - the eye is notoriously adept
at picking out apparent groupings in randomly generated distributions
of points.  A commonly used statistic is the mean surface density of
companions (henceforth the MSDC), first applied to a star forming
region (in this case Taurus) by Larson (1995) and subsequently to a
number of other star forming regions (Simon 1997; Bate, Clarke and
McCaughrean 1998; Nakajima et al. 1998; Gladwin et al. 1999), although
incompleteness in some cases limits the utility of this approach.  The
MSDC is related to the two point correlation function (Peebles 1980)
but has the advantage that it is not sensitive to the choice of
average density in the surveyed region.  It is simply computed, as a
function of angular separation, by averaging the surface density of
stars in annuli of appropriate radius placed in turn on each of the
stars in the sample.

In Taurus, the MSDC can be fitted as a power law (of slope -0.6) for
stellar separations in excess of around $0.04$ pc (Larson 1995).  A
uniform stellar distribution gives rise to a flat MSDC (equal surface
densities on all scales), so this result is immediate evidence for an
inhomogeneous stellar distribution. A power law MSDC over a large
dynamic range is moreover evidence for fractal clustering, an
interpretation favoured by Larson, although the observed MSDC over the
limited dynamic range available in Taurus is also consistent with
clustering on a single scale (Bate, Clarke and McCaughrean 1998).  The
conclusion that Taurus is indeed highly inhomogeneous is readily
confirmed by visual inspection of the stellar distribution, which
clearly shows the existence of discrete groupings containing around
$15$ stars in regions of typical size $0.5-1.1$ pc (Herbig 1977; Gomez
et al. 1993). Given the velocity dispersion measured in Taurus
(e.g. Hartmann \etal 1986; Frink et al. 1997), these groups are not
bound; this velocity dispersion is however consistent with these
groups having expanded from very compact configurations over their
assumed lifetimes. Thus the existence of the Gomez groups is
consistent with (but does not prove) an origin of stars in compact
mini-clusters.

Interpretation of the MSDC in clusters, as opposed to the more diffuse
and irregular environment of Taurus, is complicated by the global
decline of surface density with radius in this case. It turns out,
however, that if the surface density declines with distance from the
cluster centre as $R^{-1}$ or less steeply, then for clusters with no
substructure, the MSDC should still be approximately flat apart from
possible edge effects (Bate, Clarke and McCaughrean 1998). This
convenient property means that in the ONC, for
example, where the surface density declines with radius approximately
as $R^{-1}$ outside the core, the flatness or otherwise of the MSDC
can still be used as a diagnostic of clustering.

The result for the ONC is that notwithstanding the
fact that the eye can arguably  pick out apparent stellar groupings,
the MSDC is essentially flat:
i.e. the stellar distribution is statistically consistent with a smoothly
declining density law with no sub-clustering. This is not to say, however,
that sub-clustering is necessarily absent. Through generation of
synthetic clusters,  
Bate, Clarke and McCaughrean showed that over a limited region of
parameter space (i.e. for mini-clusters of a few times $10^4$ A.U. in
size), it was possible to hide a substantial fraction of the stars in
mini-clusters and yet produce an MSDC consistent with that observed. 
The range of size scales that can be hidden in this way shrinks with
the membership number of the cluster, so that unless the cluster sizes are
very finely tuned, the number of stars  contained in each  needs to be quite
small (a few tens at most).

In summary then, there is no evidence for sub-clustering within the
ONC,  
although there are patterns of sub-clustering that would not be ruled
out by the observed MSDC. (Note, however, that the {\it massive} stars
do appear to be clustered in the central regions: see Section VI). 
Rough estimates suggest that this
lack of sub-clustering may not necessarily rule out sub-clustering at
birth: although Orion is generally believed to be younger than Taurus
(Kenyon and Hartmann 1995),
the higher stellar surface density means that sub-clusters would
merge and lose their identity more rapidly during the dissolution
process. Further modeling, using all the
available kinematic and spatial data for the cluster is required in
order to rule out the possibility that the ONC was
composed of an ensemble of sub-clusters at birth.

\mainsection {{I}{V} ~~ Dynamical interactions in compact clusters} 
\backup

Mini-clusters comprising N members dissolve due to point mass gravitational
interactions on a timescale that is a strong positive function of
of N (van Albada 1968; Heggie 1974).
Thus point mass gravitational effects are the main agent of dissolution
for small N systems, where a central binary can interact and eject
the majority of stars,  whereas gas expulsion
may predominate in larger N systems (see Section V). 
Compact, small N clusters, therefore,  are short-lived even
if gas expulsion is neglected: for example, a cluster of $10$ stars in a
volume of radius $0.1$ pc dissolves in less than a million years. 
This fact underlines
the difficulty of assessing the level of sub-clustering at birth in
star forming regions, inasmuch as information on the smallest scales is
rapidly erased, sometimes before the cluster becomes optically visible.

Cluster dissolution by point mass dynamics results from the formation of
a central binary which absorbs the potential energy of the cluster, thereby
unbinding the other members. There is an overwhelming tendency for the
two most massive stars to constitute the binary (van Albada 1968). Thus
{\it if} binaries form from small N, non-hierarchical ensembles, their
pairing statistics are well defined (McDonald and Clarke 1993):  
the binary 
fraction is a strongly increasing function of primary mass, and,
unless the membership number of the mini-cluster is very small (3 or so),
there is a strong tendency for stars to pair with companions of almost equal
mass. McDonald and Clarke showed that a hallmark of binaries formed
dynamically in such small clusters is that the mass distribution
of secondaries does not depend on the primary's mass.
This property can be tested for in
binary samples with primaries of various masses.
It is clear, however, (from the fact that most solar type stars are binary
primaries, whereas most OB binaries have high mass secondaries) that this
process cannot simultaneously account for both low and high mass
binary statistics,
unless the IMF is spatially variable.

In reality, of course, one would not expect interactions in such mini-
clusters to result  purely from point mass gravity. For few body clusters,
the expected radii of circumstellar discs are a significant fraction
of the mean interstellar separation (Pringle 1989; Clarke and Pringle 1991)
so that hydrodynamic encounters with disc gas are to be expected at closest
approach (Larson 1990; Heller 1993;  
Hall, Clarke and Pringle 1996). 
Whereas the higher velocity dispersion in large N clusters renders
most such encounters disc destroying (rather than binary producing;
Clarke and Pringle 1991), the relatively slow encounters within small
N mini-clusters can lead to a substantial binary fraction through
star-disc capture (McDonald and Clarke 1995). If star-disc capture is
the dominant binary production route, the dependence of binary fraction
on primary mass is somewhat reduced, whilst the companion mass distribution
reflects almost random pairing from the IMF.

In addition to the possible production of binaries, close encounters
in mini-clusters can have two further effects. The first is the destructive
effect of star-disc encounters. It has been argued for example
(Mottmann 1977), that
the Sun may have originated in a cluster, so that 
episodes of intense meteoritic bombardment, as evidenced by the cratering record
of the terrestrial planets, would have followed perturbations to the Oort cloud
by stellar encounters.
Simulations of star-disc encounters indicate that discs are truncated at
about one third of the stars' closest approach (Clarke and Pringle
1993), 
the pruned remnant being left with an exponential radial density
profile (Hall 1997) similar to those observed in the `silhouette discs'
in Orion (McCaughrean and O'Dell 1996). Such pruning would not only
reduce the strength of disc emission (by reducing the mass and surface
area of the disc), but would also shorten the disc lifetime (mainly due
to the reduction in the disc's radial extent). It  has been noted
(e.g. Bouvier, Forestini and Allain 1997; Armitage 1996) that a wide range of disc lifetimes
are necessary both to explain the co-existence of Classical and Weak Line
T Tauri stars in the same region of the HR diagram and to explain the
spread in rotation rates of stars on the ZAMS. 

The velocities acquired by stars during the dissolution of small
clusters is of order the velocity at pericentre during a three-body
encounter. Thus whilst the majority of stars drift apart with a
velocity that exceeds the cluster escape velocity by a factor of order
unity, stars can be ejected from particularly close encounters with
considerably larger velocities. Sterzik and Durisen (1995) have
applied this model to the production of the dispersed population of
Xray sources detected by ROSAT in the vicinity of star forming regions
(Alcala et al. 1996; Neuhauser 1997), arguing that these sources are
Weak Line T Tauri stars that were formed in the smaller volume
currently occupied by the emission line (Classical T Tauri) stars, but
were ejected by dynamical encounters in small clusters (see Feigelson
1996 for an alternative view). The combination of the apparent
distance of these stars from their putative birthplaces and their ages
derived from the HR diagram implies ejection velocities greater than
$\sim 3 {\rm km}/{\rm s}$, which requires mini-clusters comprising a few
(i.e. $5-10$) stars within a radius of $500-1000$ AU.  The close
encounters (pericentre of about 0.5 A.U.) that are required to
generate such velocities shave the discs to such small radii that the
disc depletion timescale is considerably reduced. In the case of discs
that are magnetically disrupted in their innermost regions, such tidal
pruning in close encounters can lead to the system appearing as a Weak
Line T Tauri star even at the young age ($\sim 10^6$ years) inferred
for the dispersed population of Xray sources (Armitage and Clarke
1997). It should be noted, however, that many of the dispersed Xray
sources may be somewhat older foreground stars - see discussion by
Briceno et al. 1997; and Wichmann et al. 1997 - and that proper motion
data supports the ejection hypothesis only in some cases
(Neuhauser et al. 1998; Frink et al. 1997).
Clearly the controversial  question
of what proportion of the ROSAT sources are indeed runaway T Tauri
stars needs to be
settled before one can assess the required ejection rates of T Tauri
stars from star forming regions and hence the number of compact mini-clusters
that are needed to generate this ejection rate.

 In summary, then, a number of physical processes occurring in very compact
mini-clusters can profoundly affect the properties of the stars and their
associated discs. These physical processes rely on small interstellar
separations and relatively low velocity dispersions and their role is
thus negligible if estimated using the densities and velocity dispersions
of large scale star forming regions (such as, for example, the Orion
Nebula Cluster or the central regions of Taurus). If  the stars in
these regions were not considerably sub-clustered at birth, then close
encounters would have played an insignificant role and stars would have
evolved essentially independently. On the other hand, if stars were
tightly clustered at birth, then it may provide solutions to a number
of problems (e.g. that of binary formation, of the apparently large
dispersion in disc lifetimes or of the generation of runaway T Tauri stars).

\mainsection{{V}~The role of gas in clusters}
\backup

As discussed in Section~II, young stellar clusters are commonly
associated with massive cores of molecular gas (e.g. Lada~1992; Lada,
Evans and Falgarone~1997). This gas comprises the majority of the
cluster mass in the youngest systems (typically 50 to 90 \% of their
total mass, Lada~1991),  but appears to be absent in older systems (eg
IC~348 at $\approx 5 \times 10^6$ years, Lada and Lada~1995).

In addition to being a major contributor to the gravitational
potential -- and hence, by its removal, providing an obvious way to
unbind the cluster -- the gas can also interact with and be accreted
by the stars.  As pointed out by several authors (e.g.
Zinnecker~1982; Larson~1992),   accretion in a clustered environment 
may play an important role in shaping the 
observed spectrum, and segregation, of
stellar masses. Bonnell
\etal~(1997) used SPH/accretion particle simulations to study the
evolution of clusters initially comprising a few (point mass) stars
plus a distributed gas component. The stars excite gravitational wakes
in the surrounding gas (cf.  Gorti and Bhatt~1996)
and gain mass by accretion (in
these calculations no gas expulsion is included, so all the gas
ultimately ends up on the stars). 

The competitive
accretion of gas by the various stars leads to an IMF in which the
dynamic range of final stellar masses is large, even when the masses
of the initial protostellar seeds are all set to the same value. The
chief determinant of ultimate stellar mass is in this case the initial
position of the protostellar seed in the cluster potential: seeds
initially deep in the potential well acquire accreted mass rapidly from
the start, and then become hard to nudge from their central position
owing to their large masses. Seeds initially at large radii,
conversely, accrete mass slowly; being low mass objects, they are more
likely to be flung out of the cluster due to interactions with more
massive stars, and thus stop accreting altogether. Thus the interplay
of hydrodynamic accretion and point mass gravitational interactions is
such as to enhance the initial `advantage' of seeds located near the
cluster core, and generates a large dynamic range of stellar masses
from arbitrary initial conditions.  It is notable in the context of
the mass segregation observed in clusters (see Section VI), that
competitive accretion provides a natural way of producing the most
massive stars in the cluster core, and requires no gradients in the
initial conditions.

It has also been argued (Bonnell, Bate and Zinnecker 1998) that
massive stars {\it must} form in the centre of dense clusters.
These authors
consider systems that become extremely dense (up to $10^8$ pc$^{-3}$)
as they shrink due to the effects of continuing accretion of gas.
In such high density environments, massive stars can form via collisional
build up of protostellar fragments. An episode of vigorous mass loss
is then invoked to clear the cluster of gas and cause it to
re-expand (since these effects occur on timescales of $\approx 10^4$ years,
these clusters are unlikely to be directly observable in their high
density phase). The formation of massive stars through collisions
is an attractive scenario inasmuch as it avoids the classic problem
of forming them by accretion (namely that for stars more massive than
around $10 M_{\odot}$, accretion is halted by the action of radiation
pressure on dust grains).

In reality, however, gas is lost from clusters in a variety of ways.
Massive stars ($\simgt 8 M_{\odot}$) eject gas by the action of
supernovae, photoionisation and stellar winds (Whitworth~1979; Tenorio-Tagle
\etal~1986; Franco, Shore and Tenorio-Tagle~1994). It is also becoming
increasingly apparent that low mass stars can provide effective
feedback of energy into the surrounding medium through the action of
energetic molecular outflows (Eisloeffel \etal, this volume). (Note
that whereas it is difficult to sustain the case that an {\it
isolated} star can cut off its own accretion supply through the action
of outflows -- because the outflows are somewhat collimated, whereas
accretion occurs over a large solid angle, and preferentially
equatorially at small radii -- it is obviously the case that a set of
randomly orientated outflow sources in a small cluster can inflict
significant damage on the residual gas).

The fate of a particular cluster in response to gas loss depends on the 
initial gas fraction, the
removal timescale and the stellar velocity dispersion when the gas is
dispersed (Lada, Margulis and Dearborne~1984; Pinto 1987; Verschueren
and David 1989; Goodwin~1997b; see also Chapter by Elmegreen et al., this volume). If the gas comprises a
significant fraction of the total mass ($\simgreat 50 \%$) and
is removed quickly compared
to the cluster crossing time,  then the dramatic reduction in the
binding energy, without affecting the stellar kinetic energy, results
in an unbound cluster.
Alternatively, if the gas is removed over several crossing times, then
the cluster can adapt to the new potential and can survive with a
significant fraction of its initial stars. For example, clusters with
gas fractions as high as 80 $\%$ can survive with approximately
half of the stars if the gas removal occurs over 4 or more crossing
times (Lada \etal~1984).

The number and age distribution of Galactic clusters suggests that
only a few percent of all Galactic field stars can have originated in
bound clusters (Wielen 1971). However, the frequency
of cluster-mode star formation (see Section~II) and the properties 
of Galactic field binaries
(Kroupa 1995) indicate that most stars may form in clusters. The
implication is that the life-time of most young clusters
is short 
$\simless10^7$~yr (Battinelli and Capuzzo-Dolcetta 1991), which is
a natural consequence of rapid gas expulsion and low local star
formation efficiency.

\mainsection{{V}{I}.~~The initial  mass distributions and shapes of clusters}
\backup

\subsection{A.~~Mass Segregation}
%
%

A common observational finding in clusters is that the most massive
stars tend to be concentrated in the central core (e.g. Mon R2:
Carpenter \etal~1997; ONC: Hillenbrand and Hartmann~1998; NGC 6231:
Raboud and Mermilliod~1998; NGC 2157 in the LMC: Fischer \etal~1998;
SL666 and NGC 2098 in the LMC: Kontizas \etal~1998).  This mass
segregation is present even in the youngest clusters, suggesting that
it represents the initial conditions of the cluster and is not due to
its subsequent evolution. Order of magnitude arguments support this
view: the timescale for mass segregation from two-body interactions
(which drive the stellar kinetic energies towards equipartition and thus
allow the massive stars to sink to the centre) is approximately the
relaxation time (Binney and Tremaine~1987; Bonnell and Davies~1998),
which is typically very long (many crossing times) compared to the age
of the cluster. However, the segregation timescale is inversely
proportional to the stellar mass so that the most massive stars will
segregate significantly faster than this.  It is not therefore clear
{\it a priori} whether the presence of massive subsystems in the cores
of clusters (such as the `Trapezium' of OB stars in the ONC) is
attributable to dynamical effects or segregation at birth.

Bonnell and Davies~(1998) investigated this issue through  N-body simulations of stellar
clusters, exploring the timescale for  massive stars
to sink to the cluster centre as a function of their initial
location. Comparing these results with  recent
observations of the ONC (Hillenbrand~1997; Hillenbrand and Hartmann~1998)
shows that the location of the massive stars (and, in particular,
the existence of the Trapezium) {\it cannot}  be accounted
for by dynamical mass segregation but must reflect the initial conditions.
The clearest indication of this result comes from 
repeated simulations based on different random realisations
of the initial conditions. It was found that
Trapezium-like
systems were generated  with significant frequency {\sl only} if
the massive stars were initially rather centrally condensed
(i.e. within the innermost 10 \% of the stars for a 70\% probability of
Trapezium formation, or within 20 \% for a 10\% probability. (Note that
these simulations did not include gas; it is possible in principle for the
observed ONC to have expanded due to previous gas loss, in which case the
shorter dynamical timescales in its initially denser configuration may have
permitted more effective mass segregation). 

As initially discussed by Zinnecker \etal~(1993; see also Bonnell
\etal~1998), simple Jeans type arguments do not lead to the expectation
that the most massive stars should form in the centre of
dense clusters. Since these regions have high densities,  the associated
Jeans mass is {\it low},  unless the local temperature is anomalously
high. Evolutionary effects, involving accretion and protostellar collisions,
are probably required to build up massive stars in cluster cores
(see Section V). 

\subsection{B.~~Cluster morphology}
%
%

Simulations of cluster dynamics are often undertaken in spherical
geometry, motivated in part by the shapes of  globular clusters in the Galaxy.
It is however well known that some clusters are significantly flattened, 
the best studied examples being the globular clusters in the LMC.
Since some of these systems are both young (with ages, at less than
$20$ Myr, of order 10 crossing times) and significantly elliptical
(projected axis ratio on the sky $\simgt  0.7$ ), 
it would seem likely that they would have originated from aspherical
initial conditions.  Analysis of the  projected axis ratio distribution 
in these  clusters suggests that their intrinsic shapes  
are triaxial (Han and Ryden 1994), indicating that velocity anisotropy,
rather than rotational flattening, is responsible.
Further examples of flattened young clusters are found among the
SSCs (`super star clusters') 
that are conspicuous in images of some interacting galaxies 
(O'Connell et al. 1994). Here the strongly disturbed gas flows that are to be
expected in galactic encounters make cluster formation from cloud-cloud
collisions an attractive possibility (Murray and Lin 1992,
Kimura and Tosa 1996), so that flattened clusters are
a natural expectation in these environments.
In the Galaxy,  obvious examples of  flattened young clusters are 
the ONC, MonR2, and NGC~2024, where
isophotal fitting of the outer regions yields a projected axis ratio
of about 2:1 (Hillenbrand and Hartmann 1998, Carpenter et al. 1997,
Lada \etal 1991).

The interest in examining the shapes of young clusters derives from
the clues that these might give as to the mechanism for  cluster
formation.  Indeed, it is hard to think of an external trigger for cluster
formation - be it cloud-cloud collisions or the sweeping up of gas by
supernova blast waves or powerful stellar winds - that does not induce
star formation in sheet/slab-like geometry. At first sight, it might
appear most obvious to examine the shapes of the youngest embedded clusters 
in nearby
star forming regions, which are still 
associated with molecular material (see Table 1). This exercise is however
complicated by the problem of patchy extinction, plus the
difficulty of isophotal fitting in clusters
comprising  relatively few stars. Therefore the young
globular clusters in the LMC are the best laboratories for
studying this problem, since they are relatively populous and devoid of gas.

Clusters in which star formation is externally 
triggered, e.g. by a shock wave, are however unlikely to form
in virial equilibrium, so that even the youngest of the LMC globular
clusters would already have undergone a phase of violent relaxation.
Numerical simulations are therefore required in order to relate the
morphologies of observed clusters to the initial (i.e. pre-violent
relaxation) configuration of the star forming gas. This exercise 
(Boily, Clarke and Murray 1999; see also Aarseth and Binney 1978;
Goodwin 1997) yields the
answer that apart from the thinnest initial configurations (i.e. sheets
of scale height less than the mean interstellar separation, which are subject
to two-body scattering on a dynamical timescale) the system retains a
strong memory of its initial geometry during the violent relaxation
process. The relation between `initial' and `final' (i.e. relaxed) 
morphologies is set by the principle of adiabatic invariance, and yields
the prediction that the initial geometry is substantially more flattened
than that of the relaxed cluster.
When applied to the LMC globulars, initial conditions that are flattened
in the  
ratio of about 1:5 are required.

Although the degree of flattening that is required is quite substantial,
it can be generated by gas swept up in shocks of relatively low Mach
number. Since the density contrast induced in strong
shocks is of order the square of the Mach number, one sees that far flatter
configurations (axis ratio of order $10^{-4}$) would be produced, for
example, by colliding cold, thermally supported {\it homogeneous} clouds 
at relative
velocities typical of the LMC. In the case of collisions between
inhomogeneous  clouds,
density peaks carry momentum across the
net symmetry plane and generate a buckled, and thus effectively,
thicker, geometry.  The initial  morphologies deduced for the
LMC globulars may thus be compatible with
externally 
triggered cluster formation in clouds containing substantial pre-existing
density structure.

\mainsection{{V}{I}{I}.~~Theoretical considerations}
\backup

In this Section we we lay out a  very idealised conceptual
framework for cluster formation and indicate where recent theoretical
work can be slotted into this framework.

 In order to keep an open mind as to whether cluster formation is
primarily a bottom-up or top-down process, we set up a general scenario
in which the cluster progenitor gas, mass $M_{clus}$, 
consists prior to cluster formation
of an ensemble of dense lumps, mass $M_J$. Since
molecular clouds are hierarchically structured, we define  
the mass scale $M_J$ as being the
mass of {\it thermally} supported lumps that are marginally Jeans 
stable. Substructure within such lumps is not gravitationally bound,
whereas larger scale structures are supported by superthermal random
motions. We now suppose that some external trigger  over-runs the
proto-cluster region, 
which destabilises 
lumps of mass $M_J$. Each lump then collapses to form a member
of the eventual cluster (e.g. Klessen \etal~1998).
If this destabilisation promotes sub-fragmentation of the
lumps, down to a mass scale $M_*$, then the initial state of
the cluster is one of an ensemble of mini-clusters (mass $M_J$). 

Stated in this general manner, one can consider cluster formation
as occupying some position on a spectrum of possibilities. The
extreme positions are top-down fragmentation (as envisaged, for
example, in many models for globular cluster formation, e.g.
Fall and Rees 1979; Murray and Lin 1989), in
which case $M_J=M_{clus}$, and bottom-up scenarios, in which case
$M_J=M_*$. We note that top-down fragmentation engenders structures
that are coeval (to within a crossing time), whereas the age spread in
bottom-up scenarios depends on the timescale on which discrete lumps
are destabilised, and is affected, for example, by the speed with
which an external trigger over-runs the region. 

 Before proceeding further,  we here introduce
some numbers that will motivate the following discussion. Hierarchical
structures in molecular clouds obey a mass-radius (`Larson')
relation of the form $M \propto R^2$, which corresponds  
to a hierarchy of self-gravitating structures that share the same
kinetic pressure. As one descends such a hierarchy, structures of
increasing density are characterised by a decreasing velocity dispersion,
until eventually the scale is reached at which this velocity dispersion
becomes subthermal. This scale represents the minimum mass
of a self-gravitating structure within a cloud of given 
kinetic pressure (or, equivalently, $M/R^2$ for the parent GMC) and
temperature, and is thus equal to $M_J$ in the above nomenclature. Employing
canonical values for the temperature and mass-radius relation in GMCs
(respectively $T=10 K$ and $(M/1M_{\odot}) \sim (R/0.1 pc)^2$; Chieze 1987) 
one finds that $M_J$ is of order of a solar mass.

Thermally supported, self-gravitating clumps of around solar mass are
indeed observed, in nearby star forming clouds such as Taurus, as the dense cores traced by NH$_3$ (Benson and Myers~1989). The low masses of these cores implies that one
would expect top-down fragmentation to be operative only in the generation
of mini-clusters - i.e. those comprising a small number of stars. 
More populous clusters must result from a bottom-up process, i.e.
the coordinated collapse of a
number of such units. Cores that are currently forming clusters (such as 
those in
Orion) have superthermal line widths and  are thus  
presumed to be supported by
Alfvenic turbulence (Harju, Walmsley and Wouterloot~1993). 
Myers (1998) has however suggested that these cores should contain
pockets of thermally supported gas from which Alfvenic turbulence is excluded,
arguing that regions can decouple from the turbulence on size-scales 
less than the  minimum turbulent wavelength   
(this being set by the requirement that the  inverse
frequency equals the ion-neutral collision time). We will return below to
the issue of how  such thermally supported pockets might be destabilised.
 
If one considers instead the
environment in which the Galactic globular clusters
would have formed, with kinetic pressures characteristic of the proto-galaxy and
temperatures of $10^4 K$ (this marking the steep decline of the
cooling function for primordial gas), one obtains a mass scale $M_J$ of
around $10^5-10^6 M_{\odot}$ (Fall and Rees 1979). This mass is comparable 
to that of globular
clusters, suggesting that star formation in globular clusters
may well have been a top-down process. 

The issue of hierarchical
fragmentation in the top-down collapse of Jeans unstable gas has however a 
controversial history (see for example Hoyle 1953; Hunter 1962;
Layzer 1963 for
early analytical arguments for and against opacity limited
fragmentation). Larson (1978) studied the problem numerically using
a crude Lagrangian hydrodynamic code and concluded that fragmentation
does not proceed down to the opacity limit, but instead reflects the
number of Jeans masses in the gas at the initiation of collapse. 

The production of clusters by top-down fragmentation thus requires
that a clump initially containing one Jeans mass makes a rapid
(i.e. less than dynamical timescale) transition, so that it
contains a large number of Jeans masses as it enters its collapse.
This reduction in Jeans mass may be achieved either via
cooling or compression, if the system remains spherically symmetric.  
Most plausible compression mechanisms however result in the system 
becoming approximately planar: the 2D Jeans mass then depends only on
the temperature and column density, so that the fragmentation of
clouds that are swept up in shocks, for example, demands that such 
shocks cool to {\it less than} the original temperature (Lubow and
Pringle 1993; Whitworth et al. 1994). In the context of globular cluster formation
at primordial epochs, it has been suggested (e.g. Palla
and Zinnecker 1987; Murray and Lin 1989)
that protogalactic shocks activate non-equilibrium
cooling (i.e. cooling by molecular hydrogen whose formation is catalysed
by a non-equilibrium concentration of electrons in rapidly cooling gas)
and that this can effectively cool proto-globular clouds from $10^4$ K to $100$ K.  

 In the context of current star forming clouds, no such dramatic cooling
is required, since $M_J$ is already in the stellar regime and thus  
sub-fragmentation, if it occurs, will only involve a small number of pieces.
Whitworth and Clarke (1997) considered the response of Jeans stable clumps
to the mildly supersonic shocks induced by clump-clump collisions, and
concluded that cooling by dust in the dense gas behind the shock 
imposes close thermal coupling between the gas and dust: whether or not
this represents a `better than isothermal' shock (as required to promote
sub-fragmentation) of course depends on the relation between the dust
and gas temperature in the unshocked clump, which is uncertain. 
  
Bottom-up cluster formation places less stringent requirements on 
the interaction between clumps and external trigger (since the trigger
only has to destabilise the clumps rather than initiate sub-fragmentation).
Whitworth et al. (1998) have shown that the densities and temperatures
of thermally supported clumps in molecular clouds place them close to,
but somewhat outside, a regime in which dust cooling can dispose of
the compressional heating generated by collapse on a free-fall time. It
is interesting to note that if the mass-radius relation for molecular
clouds was somewhat different, so that thermally supported clumps lay
within this regime, then gas would not `hang up' at this scale but
would instead collapse to a star on a free fall time. If, conversely,
thermally supported clumps lay far from this regime, then they would
be extremely hard to destabilise and the star formation rate would
be correspondingly low.  The proximity of observed dense cores to
the dust cooling regime instead allows a situation where such cores
are stable, but may be destabilised by fairly modest perturbations
(see, for example, the suggestion of Clarke and Pringle 1997 that 
cores may be destabilised by external stirring, which widens the
bandpass for cooling in optically thick lines).  Clearly, a situation
where Jeans mass clumps are fairly stable (and hence may accumulate
in a given region) but are then fairly easy to destabilise is an optimum
one for producing clusters. Considerably more work is required, however,
before the feasibility of such ideas can be established.

\vfill\eject

\vskip.05truein
\vbox{{\hbox{\centerline{\psfig{figure=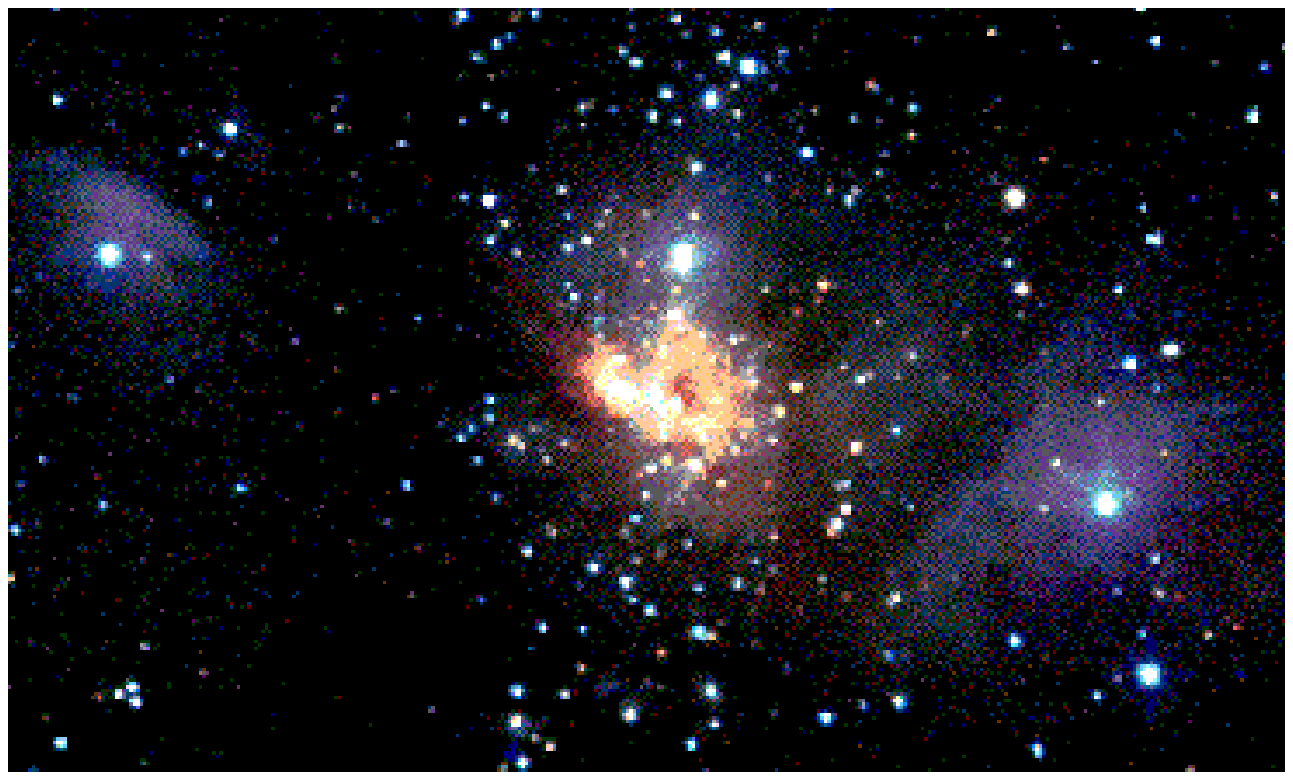,width=5.truein,height=3.truein}}}\noindent {\bf Plate 1.}  The Mon R2 cluster imaged in the near-infrared
(from Carpenter \etal~1997).  The field of view is $\sim9 \times 6$ arcmin$^2$
corresponding to $\sim2.2 \times 1.4$ pc$^2$ for a distance of 830 pc.
The bright nebulous stars near the image periphery are part of a larger chain
of reflection nebulae.   The central cluster is completely embedded
and contains $>$300 stars within a 0.4 pc diameter.
}\vskip.15truein}

\vfill\eject
\null
\vskip .5in
\centerline{\bf REFERENCES}
\vskip .25in

\ref{Aarseth, S.J., Binney, J. 1978. On the relaxation of galaxies and 
clusters from aspherical initial conditions. {\refit Mon.\ Not.\ Roy.\ Astron.\ Soc.\/} 185, 227}

\ref{Alcala, J.M., Terranegra, L. Wichmann, R. et al. 1996. New weak-line 
T Tauri stars in Orion from the ROSAT all-sky survey. {\refit Astron.\ Astrophys.\ Suppl.\/} 119,7.}

\ref{Ali, B., DePoy D.L. 1995. A 2.2 micrometer imaging survey of the Orion A 
molecular cloud. {\refit Astron.\ J.\/} 109, 709}

\ref{Ambartsumian, V.A. 1947, in {\sl Stellar Evolution and Astrophysics}, Armenian Acad. of Sci.}

\ref{Armitage, P.J. 1996. PhD Thesis, University of Cambridge.}

\ref{Armitage, P.J. and Clarke, C.J. 1997. The ejection of T Tauri stars from molecular clouds and the fate of circumstellar discs. {\refit Mon.\ Not.\ Roy.\ Astron.\ Soc.\/} 285, 540.}

\ref{Aspin, C. and Sandell, G. 1997. Near-IR imaging photometry of NGC 1333: 
a 3-mum imaging survey. {\refit Mon.\ Not.\ Roy.\ Astron.\ Soc.\/} 289, 1.}

\ref{Aspin, C. and Barsony, M. 1994. Near-IR imaging photometry of the $J-K>4$ 
sources in the Lk H$_{\alpha}$ 101 infrared cluster. {\refit Astron.\ Astrophys.\/} 288,849}

\ref{Aspin, C. and Sandell, G., and Russell, A.P.G. 1994. Near-IR imaging 
photometry of NGC 1333. I. The embedded PMS stellar population. {\refit Astron.\ Astrophys.\ Suppl.\/} 106,165}

\ref{Bate, M.R., Clarke, C.J. and McCaughrean, M. 1998. Interpreting the 
mean surface density of companions in star-forming regions. {\refit Mon.\ Not.\ Roy.\ Astron.\ Soc.\/} 297, 1163.}

\ref{Barsony, M., Kenyon, S.J., Lada, E.A., and Teuben, P.J. 1997. A 
Near-Infrared Imaging Survey of the rho Ophiuchi Cloud Core. {\refit Astrophys.\ J.\ Suppl.\/} 112, 109.}

\ref{Barsony, M., Carlstrom, J.E., Burton, M.G., Russell, A.P.G., and Garden, R. 1989, Discovery of new 2 micron sources in Rho Ophiuchi. {\refit Astrophys.\ J.\/} 346, L93}

\ref{Battinelli P., Capuzzo-Dolcetta R. 1991, Formation and evolutionary 
properties of the Galactic open cluster system. {\refit Mon.\ Not.\ Roy.\ Astron.\ Soc.\/} 249, 76}

\ref{Benson P., Myers P. 1989, A survey for dense cores in dark clouds. 
{\refit Astrophys.\ J.\ Suppl.\/} 71, 89} 

\ref{Binney J., Tremaine S. 1987, Galactic Dynamics, Princeton
        Univ. Press, Princeton}

\ref{Blaauw, A. The O associations in the solar neighborhood. 1964. 
{\refit  Ann.\ Rev.\ Astron.\ Astrophys.\/} 2, 213}

\ref{Boily, C.M., Clarke, C.J. and Murray, S.D. 1999. {\refit Mon.\ Not.\ Roy.\ Astron.\ Soc.\/} 302, 399.}

\ref{Bonnell, I.A., Bate, M.R, Clarke, C.J. and Pringle, J.E. 1997. Accretion 
and the stellar mass spectrum in small clusters. {\refit Mon.\ Not.\ Roy.\ Astron.\ Soc.\/} 285,201.}

\ref{Bonnell, I.A., Bate, M.R, Zinnecker H. 1998, On the formation of massive 
stars. {\refit Mon.\ Not.\ Roy.\ Astron.\ Soc.\/} 298, 93}

\ref{Bonnell, I.A., Davies M. B. 1998, Mass segregation in young stellar 
clusters. {\refit Mon.\ Not.\ Roy.\ Astron.\ Soc.\/} 295, 691}

\ref{Bouvier J., Forestini M., Allain S. 1997, The angular momentum evolution 
of low-mass stars. {\refit Astron.\ Astrophys.\/} 326, 1023}

\ref{Briceno, C., Hartmann, L.W., Stauffer J.R., Gagn\'e M., Stern R.A. 1997, 
X-Ray Surveys and the Post-T Tauri Problem. {\refit Astron.\ J.\/} 113, 840.} 

\ref{Carballo R., Sahu M. 1994, Near-infrared observations of new young 
stellar objects from the IRAS Point Source Catalog. {\refit Astron.\ Astrophys.\/} 289, 131.}

\ref{Carpenter, J.M., Meyer M.R., Dougados C., Strom S.E., Hillenbrand L.A. 1997, Properties of the Monoceros R2 Stellar Cluster. {\refit Astron.\ J.\/} 114, 198}

\ref{Carpenter, J.M., Snell, R.L., and Schloerb, F.P. 1995. Star Formation 
in the Gemini OB1 Molecular Cloud Complex. {\refit Astrophys.\ J.\/} 450, 201.}

\ref{Carpenter, J.M., Snell, R.L., Schloerb, F.P., and Skrutskie, M.F. 1993. 
Embedded star clusters associated with luminous IRAS point sources. 
{\refit Astrophys.\ J.\/} 407, 657.}

\ref{Chen, H. and Tokunaga, A.T. 1994. Stellar density enhancements associated 
with IRAS sources in L1641. {\refit Astrophys.\ J.\ Suppl.\/} 90,149.} 

\ref{Chieze J.-P. 1987, The fragmentation of molecular clouds. 
I - The mass-radius-velocity dispersion relations. {\refit Astron.\ Astrophys.\/} 171, 225}

\ref{Clarke, C.J. and Pringle, J.E. 1991. Star-disc interactions and binary 
star formation. {\refit Mon.\ Not.\ Roy.\ Astron.\ Soc.\/} 249,584.}

\ref{Clarke, C.J. and Pringle, J.E. 1991. The role of discs in the formation 
of binary and multiple star systems. {\refit Mon.\ Not.\ Roy.\ Astron.\ Soc.\/} 249,588.}

\ref{Clarke, C.J. and Pringle, J.E. 1993. Accretion disc response to a stellar 
fly-by. {\refit Mon.\ Not.\ Roy.\ Astron.\ Soc.\/} 261,192.}

\ref{Clarke, C.J. and Pringle, J.E. 1997. Thermal and dynamical balance in 
dense molecular cloud cores. {\refit Mon.\ Not.\ Roy.\ Astron.\ Soc.\/} 288,674}

\ref{Comeron, F., Rieke, G.H., and Rieke, M.J. 1996. Properties of Low-Mass 
Objects in NGC 2024. {\refit Astrophys.\ J.\/} 473, 294}

\ref{Comeron, F., Rieke, G.H., Burrows, A., and Rieke, M.J. 1993. The Stellar 
Population in the rho Ophiuchi Cluster. {\refit Astrophys.\ J.\/} 416, 185}

\ref{Elmegreen, B., and Lada, C. 1977, Sequential formation of subgroups in 
OB associations. {\refit Astrophys.\ J.\/} 214, 725}
     
\ref{Eiroa, C. and Casali, M.M. 1992. Near-infrared images of the Serpens cloud core - The stellar cluster. {\refit Astron.\ Astrophys.\/} 262, 468}

\ref{Feigelson, E. 1996, Dispersed T Tauri Stars and Galactic Star Formation.
{\refit Astrophys.\ J.\/} 468, 306}

\ref{Fischer P., Pryor C., Murray S., Mateo M., Richtler T. 1998, 
Mass Segregation in Young Large Magellanic Cloud Clusters. I. NGC 2157. 
{\refit Astron.\ J.\/} 115, 592}

\ref{Franco J., Shore S., Tenorio-Tagle G. 1994, On the massive star-forming 
capacity of molecular clouds. {\refit Astrophys.\ J.\/} 436, 795}

\ref{Frink, S., Roser, S., Neuhauser, R. and Sterzik, M.F. 1997. New proper 
motions of pre-main sequence stars in Taurus-Auriga. {\refit Astron.\ Astrophys.\/} 325,613.}

\ref{Gladwin P.P., Kitsionas S., Boffin H.M.J. Whitworth A.P. 1999, The structure of young star clusters. {\refit Mon.\ Not.\ Roy.\ Astron.\ Soc.\/} 302, 205}

\ref{Gorti U., Bhatt H.C. 1996, Dynamics of embedded protostar clusters 
in clouds. {\refit Mon.\ Not.\ Roy.\ Astron.\ Soc.\/} 278, 611}

\ref{Giovannetti, P., Caux, E., Nadeau, D., and Monin, J.-L.,  1998. Deep 
optical and near infrared imaging photometry of the Serpens cloud core. 
{\refit Astron.\ Astrophys.\/}. 330, 990.}

\ref{Gomez, M., Hartmann, L., Kenyon, S. and Hewett, R. 1993. On the spatial 
distribution of pre-main-sequence stars in Taurus. {\refit Astron.\ J.\/} 105,1927.}

\ref{Goodwin S.P. 1997a, The initial conditions of young globular clusters 
in the Large Magellanic Cloud. {\refit Mon.\ Not.\ Roy.\ Astron.\ Soc.\/} 286, 669}

\ref{Goodwin S.P. 1997b, Residual gas expulsion from young globular clusters. 
{\refit Mon.\ Not.\ Roy.\ Astron.\ Soc.\/} 284, 785}

\ref{Greene, T.P. and Meyer, M.R. 1995. An Infrared Spectroscopic Survey of 
the rho Ophiuchi Young Stellar Cluster: Masses and Ages from the H-R Diagram.
{\refit Astrophys.\ J.\/} 450, 233}

\ref{Greene, T.P. and Young, E.T. 1992. Near-infrared observations of young stellar objects in the Rho Ophiuchi dark cloud. {\refit Astrophys.\ J.\/} 395, 516}

\ref{Hall S.M. 1997, Circumstellar disc density profiles: a dynamic approach.
{\refit Mon.\ Not.\ Roy.\ Astron.\ Soc.\/} 287, 148}

\ref{Hall, S.M., Clarke, C.J. and Pringle, J.E. 1996. Energetics of star-disc 
encounters in the non-linear regime. {\refit Mon.\ Not.\ Roy.\ Astron.\ Soc.\/} 278,303.}

\ref{Han, C. and Ryden, B.S. 1994. A comparison of the intrinsic shapes of 
globular clusters in four different galaxies. {\refit Astrophys.\ J.\/} 433,80.}

\ref{Hanson, M.M., Howarth, I.D., and Conti, P.S. 1997. The Young Massive 
Stellar Objects of M17. {\refit Astrophys.\ J.\/} 489, 698}

\ref{Harju J., Walmsley C.M., Wouterloot J.G. 1993, Ammonia clumps in the 
Orion and Cepheus clouds. {\refit Astron.\ Astrophys.\ Suppl.\/} 98, 51}

\ref{Hartmann L., Hewwtt R., Stahler S., Mathieu R.D. 1986, Rotational and radial velocities of T Tauri stars. {\refit Astrophys.\ J.\/} 309,275}

\ref{Heller C. 1993, Encounters with protostellar disks. I - Disk tilt and 
the nonzero solar obliquity. {\refit Astrophys.\ J.\/} 408, 337}

\ref{Herbig, G.H. 1998, The Young Cluster IC 348. {\refit Astrophys.\ J.\/} 497, 736}

\ref{Herbig, G.H. and Terndrup, D.M. 1986, The Trapezium cluster of the 
Orion nebula. {\refit Astrophys.\ J.\/} 307, 609}

\ref{Heyer, M.H., Brunt, C., Snell, R.L., \etal 1996. A Massive Cometary Cloud 
Associated with IC 1805. {\refit Astrophys.\ J.\/} 464, L175.}

\ref{Hillenbrand L.A. 1995, Herbig Ae/Be Stars: An Investigation of Molecular
Environments and Associated Stellar Populations.
PhD Thesis, University of Massachusetts.}

\ref{Hillenbrand L.A. 1997, On the Stellar Population and Star-Forming 
History of the Orion Nebula Cluster. {\refit Astron.\ J.\/} 113, 1733}

\ref{Hillenbrand, L. A. and Hartmann, L.W. 1998. A Preliminary Study of the
Orion Nebula Cluster Structure and Dynamics. {\refit Astrophys.\ J.\/} 492,540.}

\ref{Heggie D.C. 1974, The role of binaries in cluster dynamics. in 
{\sl IAU Symp 62, The Stability of the Solar System and Small Stellar Systems},
ed Y. Kozai, Dordrecht, p. 225}

\ref{Hodapp, K.-W. 1994. A K' imaging survey of molecular outflow sources.
{\refit Astrophys.\ J.\ Suppl.\/} 94,615.}

\ref{Hodapp, K.-W. and Deane, J. 1993. Star formation in the L1641 North 
cluster. {\refit Astrophys.\ J.\ Suppl.\/} 88,119.}

\ref{Hodapp, K.-W. and Rayner, J. 1991. The S106 star-forming region.
{\refit Astron.\ J.\/} 102,1108.}

\ref{Hoyle, F. 1953. On the fragmentation of gas clouds into galaxies and 
stars. {\refit Astrophys.\ J.\/} 118,513.}

\ref{Hunter, C. 1962. The instability of the collapse of a self-gravitating 
gas cloud. {\refit Astrophys.\ J.\/} 136,594.}

\ref{Itoh, Y., Tamura, M., Gatley, I. 1996. A Near-Infrared Survey of the 
Taurus Molecular Cloud: Near-Infrared Luminosity Function. {\refit Astrophys.\ J.\/} 465,
L129}

\ref{Janes, K.A., Tilley, C., and Lynga, G. 1988. Properties of the open 
cluster system. {\refit Astron.\ J.\/} 95, 771}

\ref{Jones, T.J., Mergen, J., Odewahn, S., Gehrz, R.D., Gatley, I.,
Merrill, K.M., Probst, R., and Woodward, C.E. 1994. A near-infrared survey of 
the OMC2 region. {\refit Astron.\ J.\/} 107, 2120}

\ref{Kenyon, S. and Hartmann, L. 1995, Pre-Main-Sequence Evolution in the 
Taurus-Auriga Molecular Cloud. {\refit Astrophys.\ J.\ Suppl.\/} 101, 117}

\ref{Kimura, T. and Tosa, M. 1996. Collision of clumpy molecular clouds.. 
{\refit Astron.\ Astrophys.\/} 308, 979.}

\ref{Klessen, R., Burkert, A., Bate, M. 1998. Fragmentation of Molecular 
Clouds: The Initial Phase of a Stellar Cluster. {\refit Astrophys.\
J.\ Lett.\/} 501, L205}

\ref{Kontizas M., Hatzidimitriou D., Bellas-Velidis I., Gouliermis D., 
Kontizas E., Cannon R.D. 1998, Mass segregation in two young clusters in the 
Large Magellanic Cloud: SL 666 and NGC 2098. {\refit Astron.\ Astrophys.\/} 336, 503}

\ref{Kroupa P. 1995, Inverse dynamical population synthesis and star 
formation. {\refit Mon.\ Not.\ Roy.\ Astron.\ Soc.\/} 277, 1491}

\ref{Lada C. J. 1991, The Formation of Low Mass Stars: Observations. 
in {\sl The Physics of Star Formation and Early
        Stellar Evolution}, eds C. J. Lada, N. D. Kyfalis, Kluwer, p. 329}

\ref{Lada, C.J., Margulis, M., and Dearborn, D. 1984. The formation and early 
dynamical evolution of bound stellar systems. {\refit Astrophys.\ J.\/} 285, 141.}

\ref{Lada, C.J., Alves J., Lada E. A. 1996, Near-Infrared Imaging of Embedded 
Clusters: NGC 1333. {\refit Astron.\ J.\/} 111, 1964}

\ref{Lada, C.J., Young, E.T., and Greene, T.P. 1993. Infrared images of the 
young cluster NGC 2264. {\refit Astrophys.\ J.\/} 408, 471.}

\ref{Lada, E.A. 1992,Global star formation in the L1630 molecular cloud. 
{\refit Astrophys.\ J.\ Lett.\/} 393, 25L}

\ref{Lada, E.A., Depoy D. L., Evans N. J. Gatley I. 1991, A 2.2 micron survey 
in the L1630 molecular cloud. {\refit Astrophys.\ J.\/} 371 171}

\ref{Lada, E.A., Bally, J., and Stark, A.A. 1991. An unbiased survey for 
dense cores in the Lynds 1630 molecular cloud. {\refit Astrophys.\ J.\/} 368, 432.}

\ref{Lada, E.A., Evans, N.J. II, and Falgarone, E. 1997. Physical Properties 
of Molecular Cloud Cores in L1630 and Implications for Star Formation. 
{\refit Astrophys.\ J.\/} 488, 286.}

\ref{Lada E.A., Lada C.J. 1995, Near-infrared images of IC 348 and the 
luminosity functions of young embedded star clusters. {\refit Astron.\ J.\/} 109, 1682}

\ref{Larson, R.B. 1978, Calculations of three-dimensional collapse and 
fragmentation. {\refit Mon.\ Not.\ Roy.\ Astron.\ Soc.\/} 184, 69}

\ref{Larson, R.B. 1990,  Formation of star clusters. 
in {\sl Physical Processes in Fragmentation 
        and Star Formation} eds. R. Capuzzo-Dolcetta, C. Chiosi, A. DiFazio, 
        Kluwer, Dordrecht, p. 389}

\ref{Larson, R.B. 1992, Towards understanding the stellar initial mass 
function. {\refit Mon.\ Not.\ Roy.\ Astron.\ Soc.\/} 256, 641}

\ref{Larson, R.B. 1995, Star formation in groups. {\refit Mon.\ Not.\ Roy.\ Astron.\ Soc.\/} 272,213.}

\ref{Luhman, K. and Rieke, G.H. 1998. The Low-Mass Initial Mass Function in 
Young Clusters: L1495E. {\refit Astrophys.\ J.\/} 497, 354}

\ref{Li, W., Evans, N.J. II, and Lada, E.A. 1997. Looking for Distributed Star Formation in L1630: A Near-Infrared (J, H, K) Survey. {\refit Astrophys.\ J.\/} 488, 277.}

\ref{Lubow, S.H. and Pringle, J.E. 1993. The Gravitational Stability of a Compressed Slab of Gas. {\refit Mon.\ Not.\ Roy.\ Astron.\ Soc.\/} 263,701.}

\ref{McCaughrean M., O'Dell R. 1996, Direct Imaging of Circumstellar Disks 
in the Orion Nebula. {\refit Astron.\ J.\/} 111, 1977}

\ref{McDonald, J.M. and Clarke, C.J. 1993. Dynamical biasing in binary star formation - Implications for brown dwarfs in binaries. {\refit Mon.\ Not.\ Roy.\ Astron.\ Soc.\/} 262, 800.}

\ref{McDonald, J.M. and Clarke, C.J. 1995. The effect of star-disc interactions on the binary mass-ratio distribution. {\refit Mon.\ Not.\ Roy.\ Astron.\ Soc.\/} 275, 671.}

\ref{Meyer, M.R., Carpenter, J.M., Hillenbrand, L.A., and Strom, S.E. 1999. 
The Embedded Cluster Associated with NGC 2024: Near-IR
Spectroscopy and Emergent Mass Distribution. {\refit Astron.\ J.\/} submitted}

\ref{Meyer, M.R. and Lada, E.A. 1999. The Stellar Populations in the 
L1630 (Orion B) Cloud. in The Orion Complex Revisited, eds M. McCaughrean, 
A. Burkert. in press.}

\ref{Mottmann J. 1977. The Origin of the Late Heavy Bombard of Moon, Mars 
and Mercury. Icarus 31,412,}

\ref{Murray, S.D. and Lin, D.N.C. 1989. The fragmentation of proto-globular 
clusters. I - Thermalinstabilities. {\refit Astrophys.\ J.\/} 339, 933.}

\ref{Murray, S.D. and Lin, D.N.C. 1992. Globular cluster formation - 
The fossil record. {\refit Astrophys.\ J.\/} 400,265.}

\ref{Myers P. 1998, Cluster-forming Molecular Cloud Cores. {\refit Astrophys.\ J.\/} 496, L109}

\ref{Nakajima, Y., Tachihara, K., Hanawa, T. and Nakano, M. 1998, Clustering 
of Pre-Main-Sequence Stars in the Orion, Ophiuchus, Chamaeleon, Vela, and 
Lupus Star-forming Regions. {\refit Astrophys.\ J.\/} 497, 721}

\ref{Neuhauser, R. 1997. Low-mass pre-main sequence stars and their X-ray
emission. Science, 276,1363}

\ref{Neuhauser, R., Wolk, S.J., Torres, G. et al. 1998. Optical and X-ray monitoring, Doppler imaging, and space motion of the young star Par 1724 in Orion.
{\refit Astron.\ Astrophys.\/} 334, 873}

\ref{O'Connell, R.W., Gallagher,J.S. and Hunter, D.A. 1994. Hubble Space 
Telescope imaging of super-star clusters in NGC 1569 and NGC 1705. {\refit Astrophys.\ J.\/} 433,65}

\ref{Okumura, S.K., Makino, J., Ebisuzaki, T. et al. 1993. Highly Parallelized 
Special-Purpose Computer, GRAPE-3. PASJ, 45, 329}

\ref{Palla F., Zinnecker H. 1987, Non-equilibrium cooling of a hot primordial gas cloud. in {\sl Starbursts and Galaxy Evolution}, eds. T.X. Thuan, T. Montmerle, J. Tran Thanh Van,  p. 533}

\ref{Patel, N.A., Goldsmith, P.F., Heyer, M.H., Snell, R.L., and Pratap, P. 
1998, Origin and Evolution of the Cepheus Bubble. {\refit Astrophys.\ J.\/} 507, 241}

\ref{Peebles, P.J.E. 1980. The Large Scale Structure of the Universe. Princeton
University Press, Princeton, p. 138-256}

\ref{Phelps, R.L. and Janes, K. 1994. Young open clusters as probes of the 
star formation process. 1: An atlas of open cluster photometry. {\refit Astrophys.\ J.\ Suppl.\/} 90, 31}

\ref{Phelps, R.L. and Lada, E.A. 1997. Spatial Distribution of Embedded 
Clusters in the Rosette Molecular Cloud: Implications for Cluster Formation. 
{\refit Astrophys.\ J.\/} 477, 176.}

\ref{Piche F. 1993, A Near-Infrared Survey of the Star Forming Region   
NGC 2264. PASP, 105,324}

\ref{Pinto F. 1987, Bound star clusters from gas clouds with low star 
formation efficiency. PASP, 99, 1161}

\ref{Pringle J.E. 1989, On the formation of binary stars. {\refit Mon.\ Not.\ Roy.\ Astron.\ Soc.\/} 239, 631}

\ref{Raboud D., Mermilliod J.C. 1998, Evolution of mass segregation in 
open clusters: some observational evidences. {\refit Astron.\ Astrophys.\/} 333, 897}

\ref{Rieke, G.H., Ashok, N.M., and Boyle, R.P. 1989, The initial mass function
in the Rho Ophiuchi cluster. {\refit Astrophys.\ J.\/} 339, L71}

\ref{Simon, M. 1997. Clustering of Young Stars in Taurus, Ophiuchus, and the
Orion Trapezium. {\refit Astrophys.\ J.\/} 482, L81}


\ref{Sterzik, M.F. and Durisen, R.H. 1995. Escape of T Tauri stars from young 
stellar systems. {\refit Astron.\ Astrophys.\/} 304, L9}

\ref{Strom, K.M., Strom, S.E., and Merrill, M. 1993. Infrared luminosity 
functions for the young stellar population associated with the L1641 
molecular cloud. {\refit Astrophys.\ J.\/} 412, 233}

\ref{Strom, K.M., Kepner, J., and Strom, S.E. 1995. The evolutionary status 
of the stellar population in the rho Ophiuchi cloud core. {\refit Astrophys.\ J.\/} 438, 813.

\ref{Strom, S.E., \etal 1999, {\refit Astron.\ J.\/} in preparation}

\ref{Sugitani, K., Tamura, M., and Ogura, K. 1995. Young Star Clusters in 
Bright-rimmed Clouds: Small-Scale Sequential Star Formation? {\refit Astrophys.\ J.\/} 455, L39.}

\ref{Sugitani, K. and Ogura, K. 1994. A catalog of bright-rimmed clouds with 
IRAS point sources: Candidates for star formation by radiation-driven 
implosion. 2: The southern hemisphere. {\refit Astrophys.\ J.\ Suppl.\/} 92, 163.}

\ref{Sugitani, K., Fukui, Y., and Ogura, K. 1991. A catalog of bright-rimmed clouds with IRAS point sources: Candidates for star formation by radiation-driven implosion. I - The Northern Hemisphere. {\refit Astrophys.\ J.\ Suppl.\/} 77, 59.}

\ref{Testi, L., Palla, F., Prusti, T., Natta, A., and Maltagliati, S. 1997. 
A search for clustering around Herbig Ae/Be stars. {\refit Astron.\ Astrophys.\/} 320, 159}

\ref{Testi, L., Palla, F., and Natta, A. 1998. A search for clustering around Herbig Ae/Be stars. II. Atlas of the observed sources.  {\refit Astron.\ Astrophys.\ Suppl.\/} 133, 81}

\ref{Tenorio-Tagle G., Bodenheimer P., Lin D., Noriega-Crespo A.,  
        1986, On star formation in stellar systems. I - Photoionization
 	 effects in protoglobular clusters. {\refit Mon.\ Not.\ Roy.\ Astron.\ Soc.\/} 221, 635}

\ref{van Albada, T.S. 1968. The evolution os small stellar systems and the implications for double star formation. Bull. Astr. Inst. Neth., 19,479}

\ref{Verschueren W., David M. 1989, The effect of gas removal on the dynamical
evolution of young stellar clusters. {\refit Astron.\ Astrophys.\/} 219, 105}

\ref{Walter, F.M., Wolk, S.J., and Sherry, W. 1998. The $\sigma$~Orionis
    cluster. In {\sl Cool Stars, Stellar Systems, and the Sun X.}, eds.
    R. Donahue and J. Bookbinder (ASP Conf.\ Ser.\/), CD-1793.}

\ref{White, G.J., Lefloch, B., Fridlund, C.V.M., \etal 1997. An observational 
study of cometary globules near the Rosette nebula. {\refit Astron.\ Astrophys.\/} 323, 931.}

\ref{Whitmore, B.C. et al. 1998. The Luminosity Function
of Young
Star Clusters in the "Antennae" Galaxies (NGC 4038/4039). in preparation}

\ref{Whitworth A.P. 1979, The erosion and dispersal of massive molecular 
clouds by young stars. {\refit Mon.\ Not.\ Roy.\ Astron.\ Soc.\/} 186, 59}

\ref{Whitworth, A.P., Bhattal, A.S., Chapman, S.J., Disney, M.J. and Turner, 
J.A. 1994. Fragmentation of shocked interstellar gas layers. {\refit Astron.\ Astrophys.\/} 290,421}

\ref{Whitworth, A.P, Boffin, H.M.J. and Francis, N. 1998. Gas cooling by 
dust during dynamical fragmentation. {\refit Mon.\ Not.\ Roy.\ Astron.\ Soc.\/} 299,554}

\ref{Whitworth A.P. and Clarke, C.J. 1997. Cooling behind mildly supersonic 
shocks in molecular clouds. {\refit Mon.\ Not.\ Roy.\ Astron.\ Soc.\/} 291,578.} 

\ref{Wichmann, R., Krautter, J., Covino, E., et al. 1997. The T Tauri star 
population in the Lupus star forming region. {\refit Astron.\ Astrophys.\/} 320,185}

\ref{Wilking, B.A., McCaughrean, M.J., Burton, M.G., Giblin, T., Rayner, 
	J.T., and Zinnecker, H. 1997. Deep Infrared Imaging of the R Coronae 
	Australis Cloud Core. {\refit Astron.\ J.\/} 114, 2029}

\ref{Wielen R. 1971, The Age Distribution and Total Lifetimes of Galactic 
Clusters. {\refit Astron.\ Astrophys.\/} 13, 309}

\ref{Yun, J.L. and Clemens, D.P. 1994, Near-infrared imaging survey of young 
stellar objects in Bok globules. {\refit Astron.\ J.\/} 108, 612.}

\ref{Zinnecker H. 1982, Prediction of the protostellar mass spectrum in the 
Orion near-infrared cluster. in {\sl Symposium on the Orion Nebula 
        to Honour Henry Draper}, eds A. E. Glassgold \etal, New York Academy 
        of Sciences, p. 226}

\ref{Zinnecker H., McCaughrean M.J., Wilking B.A. 1993, 
The initial stellar population. in {\sl
        Protostars and Planets III}, eds. E.H. Levy, J.I. Lunine,
        Univ. of Arizona Press, p. 429}

\end

\bye